\documentclass[aps,pre,twocolumn,notitlepage,nopacs,final,letterpaper,superscriptaddress,longbibliography]{revtex4-1}

\usepackage[utf8]{inputenc}
\usepackage{amsmath,amssymb,amsthm}
\usepackage{array}
\usepackage{bm}
\usepackage{braket}
\usepackage{lipsum}
\usepackage{calc}
\usepackage{color}
\usepackage{dcolumn}
\usepackage{graphicx}
\usepackage[colorlinks=true,linkcolor=black,urlcolor=blue,citecolor=blue,anchorcolor=blue]{hyperref}
\usepackage{latexsym}
\usepackage{multirow}
\usepackage{mathtools}
\usepackage{subfigure}
\usepackage{txfonts}
\usepackage{braket}
\DeclareGraphicsExtensions{.jpg,.pdf, .mps, .png, .eps, .ps, .EPS,.gif}

\begin{document}
\def\be{\begin{equation}}
\def\ee{\end{equation}}

\def\bc{\begin{center}}
\def\ec{\end{center}}
\def\bea{\begin{eqnarray}}
\def\eea{\end{eqnarray}}
\newcommand{\avg}[1]{\langle{#1}\rangle}
\newcommand{\Avg}[1]{\left\langle{#1}\right\rangle}

\def\ie{{i.e.}\ }
\def\etal{\textit{et al.}}
\def\m{\vec{m}}
\def\G{\mathcal{G}}

\newcommand{\gin}[1]{{\bf\color{red}#1}}
\newcommand{\hanlin}[1]{{\bf\color{blue}#1}}

%\begin{document}
	\title{Higher-order percolation processes on multiplex  hypergraphs}
	\author{Hanlin Sun}
	\affiliation{School of Mathematical Sciences, Queen Mary University of London, London, E1 4NS, United Kingdom}
	\author{Ginestra Bianconi}
	\affiliation{School of Mathematical Sciences, Queen Mary University of London, London, E1 4NS, United Kingdom}
	\affiliation{The Alan Turing Institute, The British Library, 96 Euston Rd, London NW1 2DB, United Kingdom}

\begin{abstract}
Higher order interactions are increasingly recognised as a fundamental aspect of complex systems ranging from the brain to social contact networks. Hypergraph as well as simplicial complexes capture the higher-order interactions of complex systems and allow to investigate the relation between their higher-order  structure and their function.
Here we establish a general framework for assessing hypergraph robustness and we characterize the critical properties of simple and higher-order percolation processes. This general framework builds on the formulation of  the  random multiplex  hypergraph ensemble where each layer is characterized by hyperedges of given cardinality.  We reveal the relation  between higher-order percolation processes in random multiplex hypergraphs,   interdependent percolation of multiplex networks and $K$-core percolation. The structural correlations of the random multiplex hypergraphs are shown to have  a significant effect on their percolation properties. The wide range of critical behaviors observed for higher-order percolation processes on multiplex hypergraphs  elucidates the mechanisms responsible for the emergence of discontinuous transition and  uncovers  interesting  critical properties which can be applied to the study of  epidemic spreading and contagion processes on higher-order networks.
\end{abstract}

\maketitle

\section{Introduction}
Higher-order networks  \cite{battiston2020networks,torres2020and,Perspectives,courtney2016generalized,wu2015emergent}  and multilayer  networks \cite{bianconi2018multilayer,kiani2021networks,PhysReport} are generalized network structures that capture the topology of complex systems beyond the single network framework. 

Higher-order networks  include both hypergraphs and simplicial complexes and encode the set of higher-order interactions present in systems as different as social \cite{patania2017shape,st2021bursty,iacopini2019simplicial,de2020social}, ecological \cite{grilli2017higher} and brain networks \cite{giusti2016two}. Multilayer networks represent complex systems in which interactions of different nature and connotation can exist forming  networks of networks. As such multilayer networks and in particular multiplex networks are becoming the new paradigm to describe social, financial  as well as biological networks \cite{bianconi2018multilayer,kiani2021networks,PhysReport}.

Higher-order networks and multilayer networks display a very rich interplay between their structure and their dynamics \cite{bianconi2018multilayer}. Notably multilayer networks are characterized by very relevant correlations \cite{min2014network,nicosia2015measuring} that have the ability to modify the critical properties of the dynamics defined on these structures.   On their turn, higher-order networks  reveal unexpected phenomena in the context of  synchronization transitions \cite{millan2020explosive,millan2018complex,skardal2019abrupt,ghorbanchian2020higher,salova2021cluster}, diffusion \cite{torres2020simplicial,carletti2020random,millan2021local,carletti2020dynamical} and spreading processes  \cite{iacopini2019simplicial,landry2020effect,de2020social,st2021bursty,matamalas2020abrupt,jhun2019simplicial}.

In this work we investigate the interplay between structure and the dynamics of higher-order networks  providing a comprehensive multilayer framework  to study higher-order percolation processes on hypergraphs.

Percolation \cite{Doro_crit,li2021percolation,Ziff_Review,Kahng_Review} is a fundamental dynamical process defined on networks that  predicts the fraction of nodes in the giant component of a network. Having a non-zero giant component is the minimal requisite for observing collective phenomena on networks, emerging from epidemic spreading, diffusion and opinion dynamics. Therefore studying percolation of a given network has important consequences for investigating a wide range of  dynamical properties defined on networks.

Percolation theory has been extensively studied in single networks since the early days of Network Science \cite{Doro_crit,li2021percolation,Ziff_Review,Kahng_Review}.  In particular node and link percolation have been investigated in random networks with arbitrary degree distribution. In node percolation nodes are initially damaged with probability $1-p$, in link percolation links are initially damaged with probability $1-p$. In  both percolation models the fraction of nodes in the giant component is studied as a function of $p$ characterizing how the network is dismantled/disconnected by increasing the entity of the initial damage. Interestingly it has been found that the critical properties of percolation are strongly affected by the network topology of the underlying network. In particular a classic result of percolation theory is that scale-free networks are robust to random damage and in the infinite network limit can sustain a non-zero fraction of nodes in the giant component for any non-zero value of $p$ \cite{Havlin_single,newman}. 
While percolation on single networks is a continuous second order transition, $K$-core percolation \cite{dorogovtsev2006k,goltsev2006k}, studying the emergence of the $K$-core with $K\geq 2$ in complex networks can display a discontinuous hybrid transition if the degree distribution of the network has finite second moment. 

With the recent surge of interest on generalized network structures  percolation theory has further expanded thanks to   the formulation of the interdependent percolation in multiplex networks that displays a discontinuous hybrid transition in correspondence of large avalanches of failure events \cite{Havlin,Baxter,bianconi2018multilayer,son2012percolation}. In interdependent percolation the order parameter is the fraction of nodes in the mutually connected giant component that is the giant component formed by nodes connected by at least one path in each layer of the multiplex network. Interdependent percolation on multiplex networks is highly affected by the correlations \cite{min2014network,nicosia2015measuring,bianconi2013statistical} of the underlying multiplex network structure. Indeed both interlayer degree correlations \cite{min2014network} and link overlap \cite{cellai2016message,cellai2013percolation,min2015link} have been shown to have a very significant effect on the critical properties of interdependent percolation.  This field has been growing at a very fast pace and many results related to the robustness and resilience of multiplex networks have been obtained including the formulation of interdependent percolation in network of networks \cite{bianconi2014multiple}, weak percolation \cite{baxter2014weak,baxter2020exotic}, optimal percolation \cite{osat2017optimal,santoro2020optimal}, combinatorial optimization problems \cite{jerrum2021size}, $K$-core multiplex percolation \cite{azimi2014k} and percolation with redundant interdependencies \cite{radicchi2017redundant}.

This very important subject in network theory  has contributed to  a much deeper theoretical understanding of the mechanisms leading to discontinuous percolation transitions in complex networks (see recent review articles \cite{d2019explosive,boccaletti2016explosive}).

Recently the rich interplay between network geometry and topology and the critical properties of the percolation transition on higher-order networks has also gained increasing attention. Hyperbolic simplicial complexes which can be treated within the real-space renormalization group \cite{boettcher2012ordinary,bianconi2018topological,bianconi2019percolation,kryven2019renormalization,sun2020renormalization},
have been shown to reveal a rich phase diagram including discontinuous transitions for standard link percolation. Moreover homological percolation \cite{bobrowski2020homological,lee2020homological} has been show to characterize  the emergence of a non trivial homology for higher-order network topologies.

If we focus on random hypergraphs which do not display an hyperbolic network geometry, the critical properties of percolation an higher-order percolation process are just starting to be explored. 
Indeed  some interesting results related to core percolation in hypergraphs have been recently published in Ref. \cite{coutinho2020covering}. However so far the  investigation of  percolation  on random hypegraphs has been  restricted to  very simple cases of hypergraphs which hyperedges have fixed cardinality \cite{ghoshal2009random,bradde2009percolation}.

In this paper  we  relate higher-order percolation on hypergraphs  to generalized percolation processes in multiplex networks.
Random hypergraphs can have a non-trivial underlying multiplex topology  in which each layer capture the set of hyperedges of a given cardinality.  This allows us to define ensembles of random multiplex hypergraphs  in which each node is assigned set of generalized degrees $\{k_i^{[m]}\}$ where $k_i^{[m]}$ indicates the  number of hyperedges of cardinality $m$ incident to node $i$. 
As such multiplex hypergraphs are characterized by important interlayer generalized hyperdegree  correlations.
Here we show that standard percolation is affected by the non-trivial topology of multiplex hypergraphs and by their interlayer correlations that can be tuned to increase of decrease the percolation threshold of the hypergraph.
Most importantly our work reveal how the multiplex nature of the multiplex hypergraph ensembles can be exploited to propose higher-order percolation problems displaying a rich interplay between higher-order topology and dynamics and a rich set of phenomena, including discontinuous hybrid transitions and multiple percolation transitions.
  
The paper is structured as follows: in Sec II we present the random multiplex hypergraph model and we compare the model with the already widely used model of random hypergraphs; in Sec III and Sec. IV we investigate the properties of standard node and link percolation on the random hypergraphs and on the random multiplex hypegraphs respectively; in Sec. V we provide a general framework to study higher-order percolation processes on random multiplex hypergraphs; finally in Sec. VI we provide the concluding remarks.

\section{Hypergraphs models}
\subsection{Random hypergraphs}
In this paragraph we introduce random hypergraphs used widely in the literature. This model will be subsequently compared with the model of random multiplex hypergraph which allows us to capture more rich hypegraphs topologies.
Hypegraphs $\mathcal{H}=(V,H)$ are formed by a set $V$ of $N$ nodes and a set $H$ of hyperedges of different cardinality $m\leq M$. 
The number of hyperedges incident to a node is also called its {\em hyperdegree}. Therefore if all hyperedges have cardinality $m=2$, \ie all hyperedges are essentially links describing pairwise interactions,  then the hypergraph reduce to a  network, and the definition of hyperdegree reduces to the definition of degree.  In general, in hypergraphs containing hyperedges of different cardinality, the hyperdegree counts the number of hyperedges incident to a node regardless of their cardinality.

The simplest model of hypergraph here called the {\em random hypergraph model} is a maximum entropy hypergraph model with given hyperdegree  distribution $P(k)$ and distribution $\hat{P}(m)$ of hyperedge cardinalities. Therefore as far as a node has a given hyperdegree $k$ drawn from the hyperdegree distribution $P(k)$, the model is agnostic on the cardinality of its incident hyperedges.

The maximum entropy ensemble of random hypergraphs is a simple model that is strictly related to maximum entropy factor graphs. Factor graphs  are bipartite networks $G_B(V,U,E)$ formed by a set of nodes $V$ and a set of factor nodes $U$ which do not overlap and a set  $E$ of  pairwise interactions with each interaction linking a   node to a  factor node. The factor graph is simply related to the hypergraph by a simple mapping. The set $V$ of nodes of the factor graph maps to  the set $V$ nodes of the hypergraph. Each factor node of the set  $U$ is in one-to-one correspondence with the hyperedges of the hypergraph. The hyperdegree  distribution $P(k)$ and distribution $\hat{P}(m)$ of hyperedge cardinalities correspond to the degree distribution of the nodes and of the factor nodes of the factor graph respectively.

In the uncorrelated hypergraph ensemble, as long as the hyperdegree distribution and the distribution of hyperegde cardinality have a structural cutoff, the probability that node $i$ is connnected to hyperedge $\alpha$ is given by 
\bea
\tilde{p}_{i\alpha}=\frac{k_im_{\alpha}}{\avg{k}N}
\eea
where $k_i$ indicates the hyperdegree of node $i$ and $m_{\alpha}$ indicates the cardinality of the hyperedge/factor node $\alpha$.
The corresponding hypergraph includes and hyperedge $\alpha=[i_1,i_2,\ldots,i_m]$ with probability  \cite{courtney2016generalized}
\bea
p_{[i_1,i_2,\ldots,i_m]}=(m-1)!\frac{\prod_{r=1}^mk_{i_r}}{(\avg{k}N)^{m-1}}.
\eea
\begin{figure*}[hbt!]
	\includegraphics[width=1.7\columnwidth]{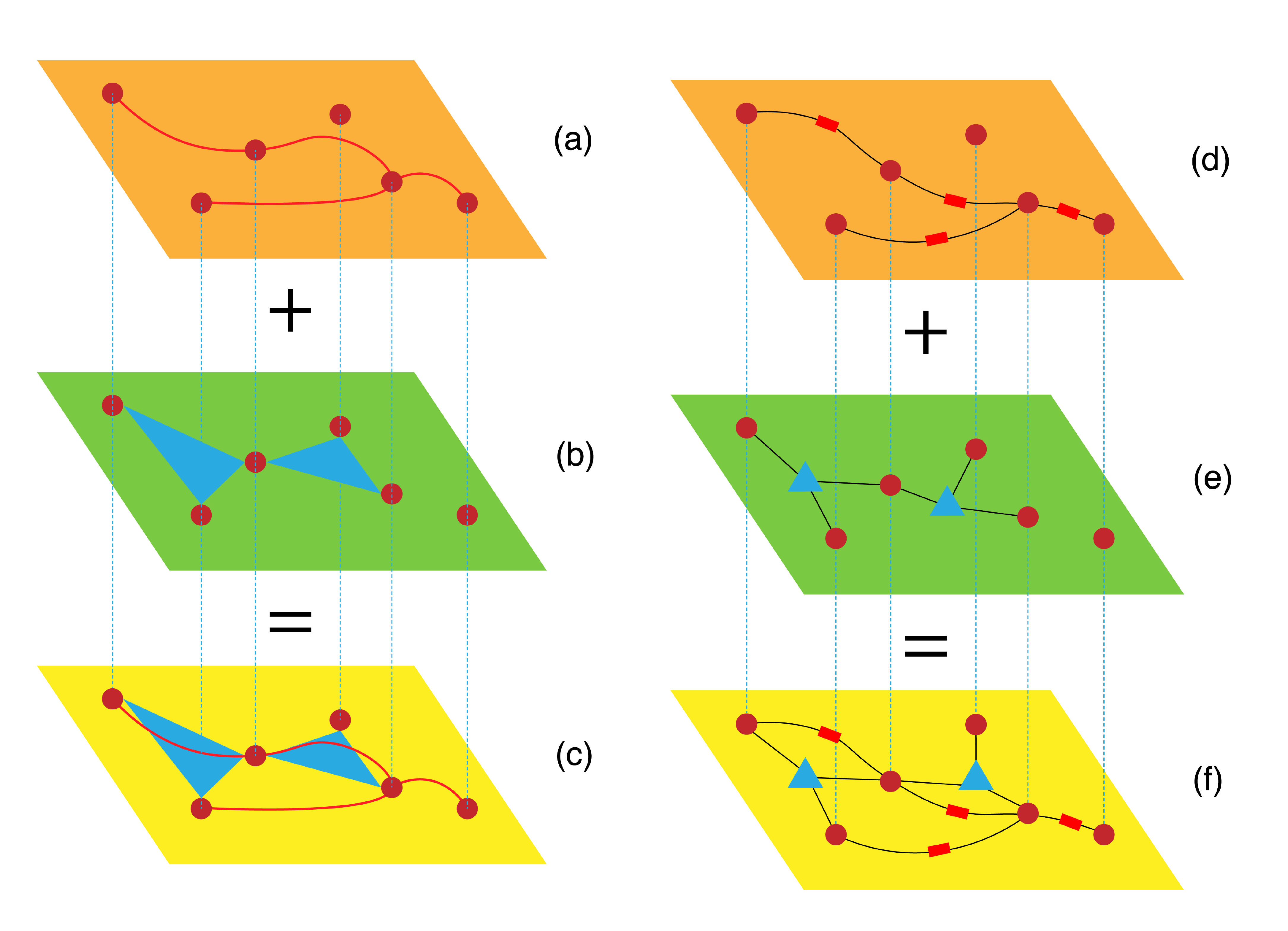}
	\caption{A schematic representation of the multiplex network construction of the hypergraph with given generalized hyperdegree sequences for hyperedges of cardinality $m_1=2$ (layer 1) and $m_2=3$ (layer 2). First a  configuration model is used to generate  a simple network capturing the $2$-body interactions of the hypergraph (panel a). Secondly  the configuration model of simplicial complexes \cite{courtney2016generalized} is used to generate a  pure simplicial complex formed exclusively by triangles. Only the information about the $3$-body interactions is retained (panel b).  Finally the information of the different layers is aggregated to generate the desired hypergraph including hyperedges of size $m=2$ and $m=3$ (panel c). This construction can be generalized to an arbitrary number of layers.  The   factor graph representation of the mulitiplex hypergraph is shown in panels (d), (e), and (f).}
	\label{fig:multiplex}
\end{figure*}
\subsection{Random multiplex hypergraphs } 

The random hypergraphs described in the previous paragraph are maximum entropy ensemble in which we fix the hyperdegree of each node.  Here we consider  hypergraphs in which we assign to each node a set of generalized hyperdegrees each one fixing the incident number of hyperedges of a given cardinality. This allow  to control the number of hyperedges of a given cardinality incident to each node and providing a more refined hypergraph model than the random hypergraph.

As we will see this model can be mapped to a multiplex network model \cite{bianconi2013statistical,bianconi2018multilayer} hence we indicate this model as {\em random multiplex hypergraph}.

In this case, as in the previous case,  we consider an ensemble of hypegraphs $\mathcal{H}=(V,H)$ formed by a set $V$ of $N$ nodes and a set $H$ of hyperedges of different cardinality $m\leq M$.  The hypergraph ${\mathcal{H}}$ in the multiplex hypegraph  ensemble is  determined by a set  tensors of dimension $2\leq m\leq M$ where the $m$-th tensor ${\bf a}^{[m]}$ determines all the hyperedges of dimension $m$, \ie,  it has elements $a^{[m]}_{i_1i_2\ldots i_m}=1$ only if $\alpha=[i_1,i_2\ldots,i_m]\in H$, otherwise $a^{[m]}_{i_1 i_2\ldots i_m}=0$.
Each node $i$ is assigned a set a set of generalized hyperdegrees 
\bea
{\bf k}_i=(k^{[2]}_i,k^{[3]}_i,\ldots k^{[M]}_i)
\eea
where $k_i^{[m]}$ indicates the number of hyperedges of degree $m$ incident node $i$, \ie,
\bea
k_i^{[m]}=\sum_{j_1,j_2,\ldots, j_{m-1}}a^{[m]}_{i,j_1,j_2\ldots, j_{m-1}}.
\eea
For these  hypergraphs  we can define    the  generalized hyperdegree distribution $P({\bf k})$  as the probability that a random node of the hypergraph has generalized hyperdegrees ${\bf k}_i={\bf k}$ with 
${\bf k}=\{k_m\}_{2\leq m\leq M}.$

The {\em random multiplex hypergraph}  is  the maximum entropy hypergraph model with given generalized hyperdegree distribution $P({\bf k})$  and given distribution $\hat{P}(m)$ of cardinality of the hyperedges.  

In the  hypergraph setting there are no   constraints relating hyperedges of different cardinality. Therefore the hyperedges of different cardinality can be drawn independently. This is different from what happens in the simplicial complex setting where if a simplex indicating interaction between $m$ nodes is present in the simplicial complex, also all the  simplices formed by the proper subset of its nodes belong to the simplicial complex.
Despite this difference, simplicial complex models can be very efficiently used to model hypergraphs.
Indeed the maximum entropy hypergraph with given generalized hyperdegree sequences can be constructed starting from the well establish configuration model of pure simplicial complexes \cite{courtney2016generalized} by mapping the hypergraph to a multiplex network \cite{de2020social} in which every layer indicates the interactions described by hyperedges of a given size $m$. 

The algorithm to construct a random multiplex hypergraph (see Figure $\ref{fig:multiplex}$) is:
\begin{itemize}
\item[(1)]
Consider a multiplex networks with $M-1$ layers $m$ with $2\leq m<M$ and $N$ nodes corresponding to the $N$ nodes of the hypergraph.
\item[(2)]
For each layer $m$ consider configuration model of pure $(m-1)$-dimensional simplicial complexes \cite{courtney2016generalized} with generalized degree sequence 
\bea
\{k^{[m]}\}_{i=1,2\ldots, N}=\{k_1^{[m]},k_2^{[m]},\ldots, k_N^{[m]}\}.
\eea  From this simplicial complex extract the hypergraph formed only by the simplicial complex facets. This hypergraph is defined by a tensor ${\bf a}^{[m]}$  describing  all the $m$-body interactions of the multiplex hypergraph.
For simplicity we assume that the hyperdegrees $\{k_i^{[m]}\}$ display a structural cutoff $\tilde{K}^{[m]}$ given by 
\bea
\tilde{K}^{[m]}=\left[\frac{(\avg{k^{[m]}N})^{m-1}}{(m-1)!}\right]^{1/m}.
\eea
In this hypothesis   the probability $p_{[i_1,i_2\ldots,i_m]}^{[m]}$
 of the hyperedge $[i_1,i_2\ldots,i_m]$ is given by \cite{courtney2016generalized}
 \bea
 p_{[i_1,i_2\ldots i_m]}^{[m]}=(m-1)!\frac{\prod_{r=1}^mk_{i_r}^{[m]}}{(\avg{k^{[m]}}N)^{m-1}}.
 \label{p}
 \eea
 Note that some layers might be empty if 
 \bea
 \sum_{i=1}^Nk^{[m]}_i=0.
 \eea
 In this case the number of layers of the random multiplex hypegraph is given by the number $M'$ of layers with at least one hyperedge.
\item[(3)]
Consider the  hypergraph obtained by aggregating all the layers, \ie considering all the interactions of different sizes $2\leq m<M$. Note that this aggregated hypergraph, differently from the aggregated multiplex network with pairwise interactions, retains its multilayer nature as the hyperedges of different cadinality can be easily distinguished also in the aggregated version of the hypergraph. Therefore we will not make a distinction between this aggregated hypergraphs and their multiplex representation.
\end{itemize}

The random multiplex hypergraph ensemble can also be viewed as a multiplex networks of factor graphs where each layer is a factor graph in which factor nodes have fixed degree. In this factor graph interpretation, the probability that node $i$ is connected to factor node $\alpha$ in a given layer describing $m$-body interactions, \ie with $m_{\alpha}=m$ is given by 
 \bea
\tilde{p}_{i\alpha}=\frac{k_i^{[m]}m}{\Avg{k^{[m]}}N}.
\eea

The mapping between the hypergraph model with given generalized hyperdegree sequences and mutliplex networks, allows us to address the role of correlations have in this hypergraph model.
Indeed, by considering a parallelism to multiplex networks we can investigate different types of possible correlations in hypergraph models.
 First of all the  hyperdegrees ${\bf k}$ of a given nodes can be correlated. In a hypergraph including $m=2$ and $m=3$ hyperedges positive generalized hyperdegree correlations  indicate for instance that  nodes with many $2$-body interactions have also many $3$-body interactions and nodes with few $2$-body interactions have also few $3$-body interactions. On the contrary negative correlations of  generalized hyperdegrees will imply that nodes with many $2$-body interactions will participate in few $3$-body interactions and vice-versa.
 Secondly we might be interested in the overlap between hyperedges of different cardinality. This implies that in a hypergraph including $m=2$ and $m=3$ hyperedges, we might be interested to assess how many $2$-body interactions connect nodes already connected in $3$ body interactions.

In this work we will focus in particular in the effect of the correlations between  generalized hyperdegrees on the robustness properties of hypergraphs. 
Indeed we notice that in the considered hypergraph ensemble hyperedges do not have a significant overlap.
To show that that we define the total overlap $O^{[m,n]}$ between $m$-hyperedges and $n$-hyperedges with $n>m$ as 
\bea
O^{[m,n]}=\sum_{\alpha\in Q_{[m]}} a_{\alpha}^{[m]}A_{\alpha}^{[n]}.
\eea
where $Q_{[m]}$ indicates the set $m$-tuples of nodes of the hypergraph   and $A_{\alpha}^{[n]}=1$ if and only if $\alpha$ is a subset of nodes of an existing $n$-hyperedges, otherwise $A_{\alpha}^{[n]}=0$.
The average overlap $\Avg{O^{[m,n]}}$ over the hypergraph ensemble with marginals given by Eq. (\ref{p}) reads 
\bea
\avg{O^{[m,n]}}=\sum_{\alpha\in Q_{[m]}}p_{\alpha}^{[m]}p_{\alpha}^{[n,m]},
\eea
where $p_{\alpha}^{[m]}$ is given by Eq. (\ref{p}) and where $p_{\alpha}^{[n,m]}=\Avg{A_{\alpha}^{[n]}}$ is given by 
\bea
 p_{[i_1,i_2\ldots i_m]}^{[n,m]}=\frac{(n-1)!}{(n-m)!}\frac{\prod_{r=1}^mk_{i_r}^{[n]}}{(\avg{k^{[n]}}N)^{m-1}}.
\eea
This implies that the average overlap $\avg{O^{[m,n]}}$ is negligible for $N\gg1 $ as it scales as 
\bea
\avg{O^{[m,n]}}=\frac{(n-1)!(m-1)!}{(n-m)!m!}\frac{\Avg{k^{[n]}k^{[m]}}^{m}}{\Avg{k^{[n]}}^{m-1}\Avg{k^{m}}^{m-1}N^{m-2}},
\eea
(see analogous treatment for multilayer networks in \cite{bianconi2013statistical}).
Therefore the overlap of hyperedges is negligible in the sparse regime where the marginals are expressed by Eq. $(\ref{p})$.

\section{Percolation on random hypergraphs}

\begin{figure}
	\includegraphics[width=\columnwidth]{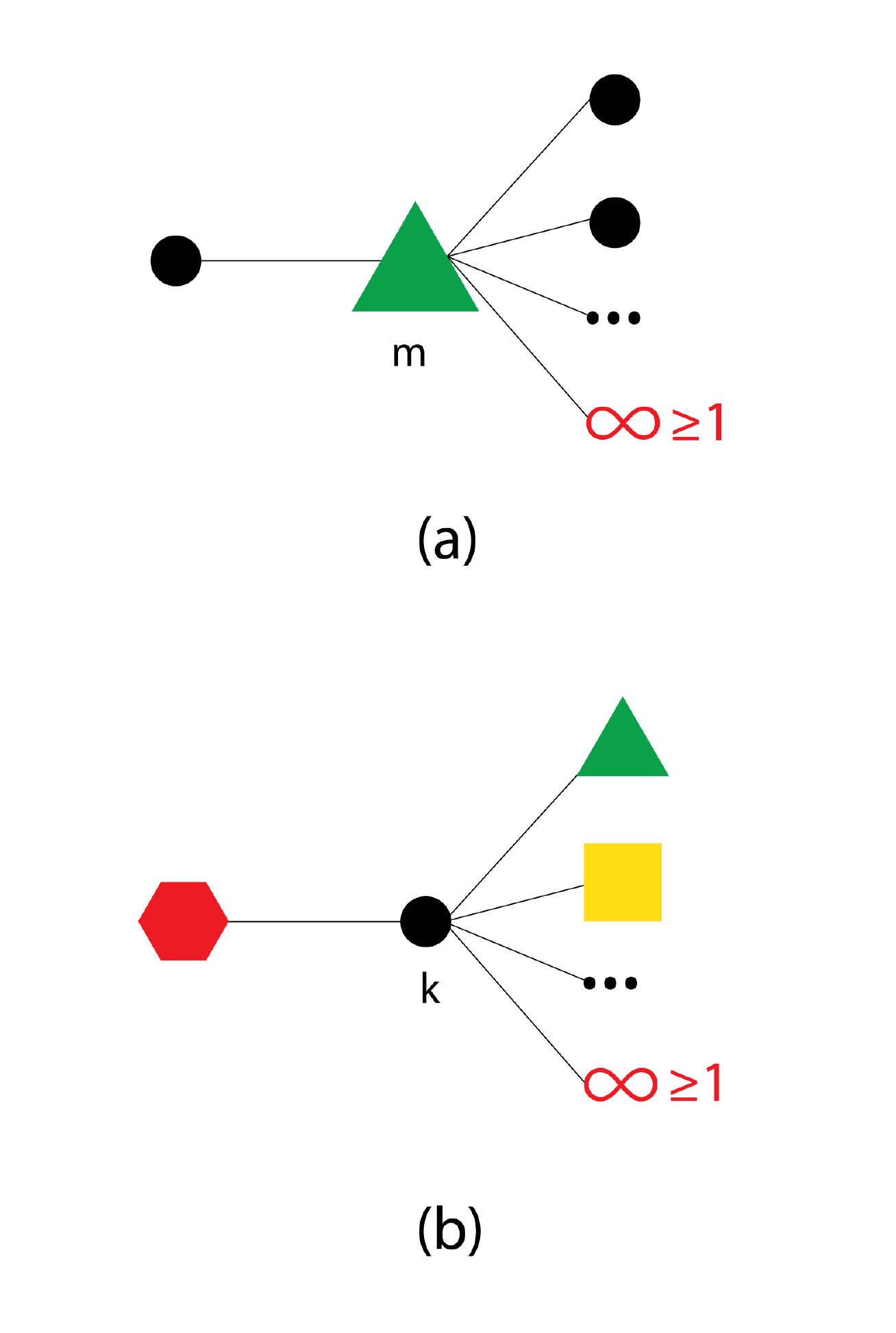}
	\caption{
An schematic  illustration of Eqs.~\eqref{eq:S_random} for $\hat{S}$ and $S$ are shown in panels (a) and (b) respectively. Black circles represent nodes, triangles,squares and hexagons represent factor nodes (hyperedges) with different cardinality. 	}
	\label{fig:diagram_random}
\end{figure}

Percolation in random hypergraphs can be treated directly by extending the ideas and concepts of percolation on factor graphs.
Therefore, since the factor graph corresponding to a random hypegraph is locally tree-like we can write self-consistent equations for the probability $\hat{S}$ that starting from a node and following a link we reach a factor node (hyperedge) in the giant component and for the probability $S$ that starting from a factor node (hyperedge)  and following a link of the factor graph we reach a node in the giant component. 
Assuming that each node is not initially damaged  with probability $p^{[N]}$ and each hyperedge is not initially damaged with probability $p^{[H]}$ the self consistent equations for $S$ and $\hat{S}$ read:
\bea
	\hat{S} &=& p^{[H]}\sum_{m} \frac{m}{\langle m \rangle} \hat{P}(m) \left[1 - (1-S)^{m-1}\right], \nonumber \\
	S &=& p^{[N]}\sum_{k}\frac{k}{\langle k \rangle}P(k)\left[1 - (1 - \hat{S})^{k-1}\right].\label{eq:S_random}
\eea
A diagramatic representations of these two equations is shown in Figure $\ref{fig:diagram_random}$.
%\end{document}

The percolation problem is fully characterized by its order parameters given by the probability $R$ of finding a node in the giant component  and the probability $\hat{R}$ of finding a hyperedge in the giant component. In a random hypergraph, these order parameters can be expressed in terms of $\hat{S}$ and $S$ as  
\bea
	R &=& p^{[N]}\left(1 - \sum_{k} P(k) (1 - \hat{S})^{k}\right),\nonumber \\
	\hat{R}& =& p^{[H]}\left(1 - \sum_{m} \hat{P}(m) (1 - S)^{m}\right).
	\label{eq:R_random}
\eea
These equations together with the self-consistent Eqs. (\ref{eq:S_random}) can be used to investigate the critical properties of percolation inferring the robustness of the random hypergraph. In particular  we can impose $p^{[H]}=1$ (or $p^{[N]}=1$) and to characterize node percolation (or hyperedge percolation)  where only hyperedges are randomly removed (or node percolation where only nodes are randomly removed). If the hypergraph only contains hyperedges of cardinality $m=2$,( i.e. it reduces to a network) these two percolation problems reduce to link and node percolation respectively.

\begin{figure}
	\includegraphics[width=\columnwidth]{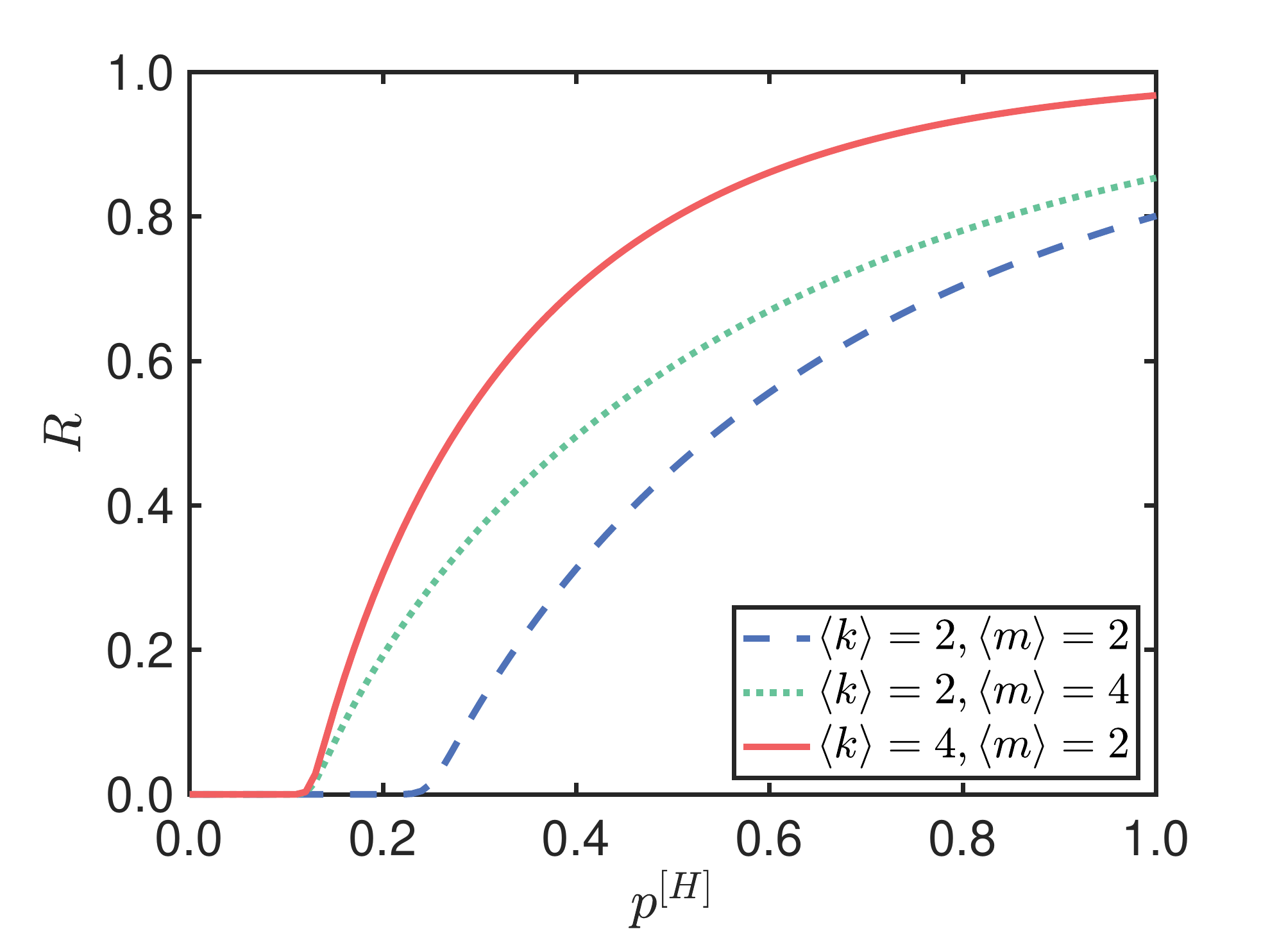}
	\caption{The fraction of  nodes in the giant component $R$ is shown versus   $p^{[H]} = p$ for random hypergraphs. The hyperdegree distribution $P(k)$ and distribution of cardinality of hyperedges $\hat{P}(m)$ are  Poisson distribution, with different  expectation $\avg{m}$ and $\avg{k}$.}
	\label{fig:random}
	\end{figure}
In Figure $\ref{fig:random}$ we show $R$ versus $p^{[H]}=p$ for hyperedge percolation ($p^{[N]}=1$) when both the hyperdegree distribution and the distribution of cardinalities of hyperedges are Poisson distributed.  
The critical point of the general percolation problem defined in Eq. (\ref{eq:R_random}) is characterized by the  critical thresholds $p_c^{[H]}$ and $p_c^{[N]}$.  By imposing that the largest  eigenvalue of the Jacobian matrix of Eqs. (\ref{eq:S_random}) is equal to one at $S=\hat{S}=0$, we find that $p_c^{[H]}$ and $p_c^{[N]}$ must satisfy 
\begin{equation}
	p_c^{[N]}p_c^{[H]} \frac{\langle k(k-1) \rangle}{\langle k \rangle} \frac{\langle m(m-1) \rangle}{\langle m \rangle} = 1.
	\label{eq:pcs_random}
\end{equation}
Therefore for hyperedge percolation in which $p^{[N]}=1$ we obtain that the critical threshold $p_c^{[H]}$ satisfies 
\begin{equation}
p_c^{[H]} \frac{\langle k(k-1) \rangle}{\langle k \rangle} \frac{\langle m(m-1) \rangle}{\langle m \rangle} = 1,
	\label{eq:pcH_random}
\end{equation}
for node percolation in which $p^{[H]}=1$ we obtain that the critical threshold $p_c^{[N]}$ satisfies 
\begin{equation}
	p_c^{[N]}\frac{\langle k(k-1) \rangle}{\langle k \rangle} \frac{\langle m(m-1) \rangle}{\langle m \rangle} = 1.
		\label{eq:pcN_random}
\end{equation}
We note that  Eqs.(\ref{eq:pcH_random})  and (\ref{eq:pcN_random}) fixing the hyperedge and the node percolation thresholds are invariant if we permute the distributions $P(k)$ and $\hat{P}(m)$. This effect can be seen also in Figure $\ref{fig:random}$ where it is evident that the hyperedge pecolation thresholds of two random hypegraphs with Poisson $P(k)$ and Poisson $\hat{P}(m)$  is the same if for the first hypergraph $\avg{k}=2$ and $\avg{m}=4$ and for the second hypergraph  $\avg{k}=4$ and $\avg{m}=2$.

Finally we note that the graph is  formed only  by $m$-hyperedges, the distribution $\hat{P}(m^{\prime})$ reduces to a $\delta$-function:
\begin{equation}
	\hat{P}(m^{\prime}) = \delta_{m^{\prime}, m},
\end{equation}
and Eq. (\ref{eq:pcs_random}) reduces to:
\begin{equation}
	p_c^{[N]}p_c^{[H]} (m-1) \frac{\langle k(k-1) \rangle}{\langle k \rangle} = 1.
	\label{eq:pc_random_m}
\end{equation}
This last equations reduces to results obtained in Refs. \cite{ghoshal2009random,bradde2009percolation}.

\section{Percolation on  random multiplex hypergraphs}
\subsection{General framework}

\begin{figure}
	\includegraphics[width=\columnwidth]{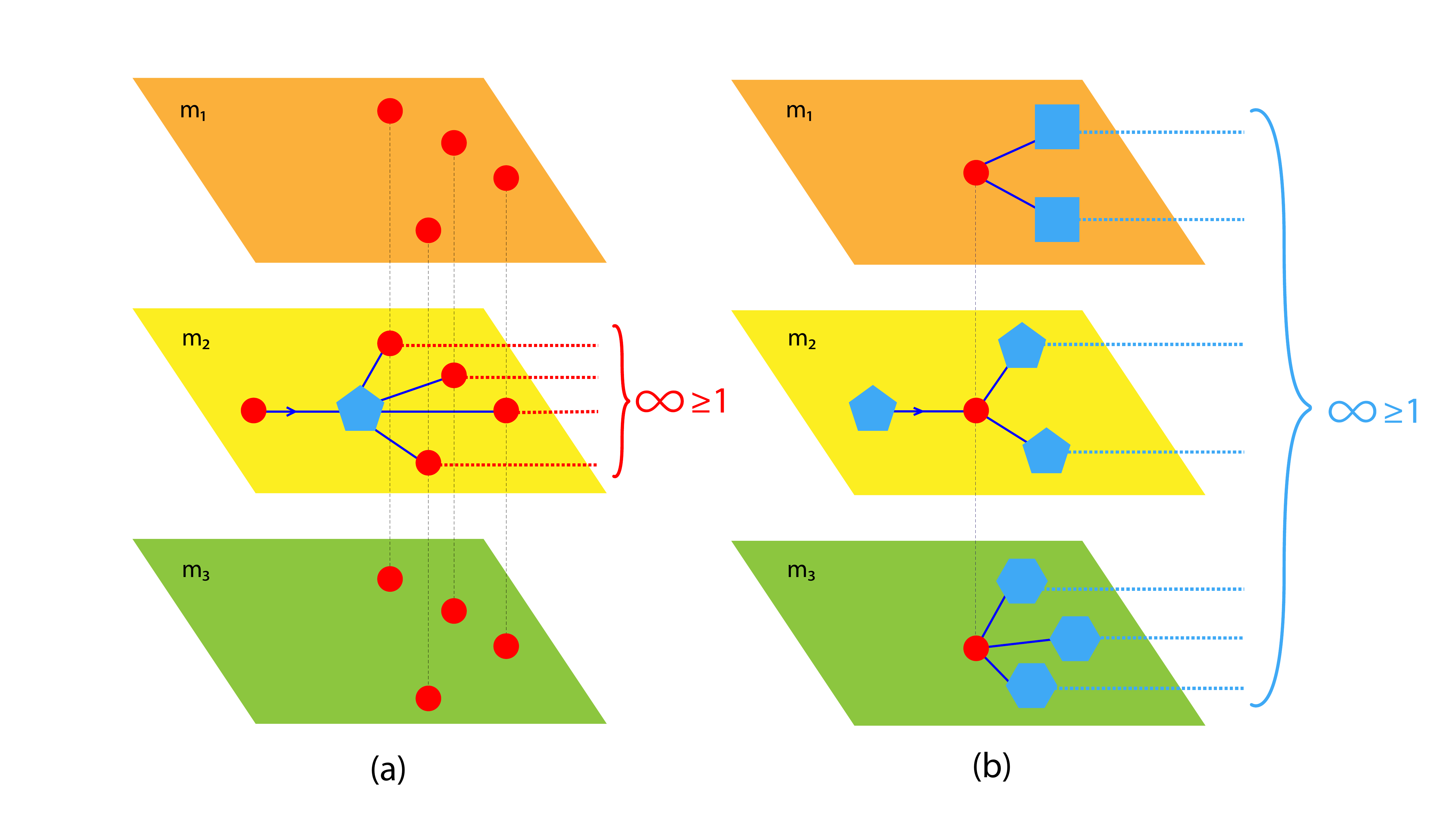}
	\caption{A schematic  illustration of Eqs.~\eqref{eq:S_multiplex} for $\hat{S}_m$ and $S_m$ are shown in panels (a) and (b) respectively. Red circles represent nodes, squares, pentagons and hexagons represent factor nodes (hyperedges) with different cardinality.  }
	\label{fig:diagram_random_multiplex}
\end{figure}

We consider percolation on  random multiplex hypergraphs, \ie   hypergraphs with given generalized hyperdegree sequences when nodes are not initially damaged with probability $p^{[N]}$ and hyperedges are not initially damaged with probability $p^{[H]}$.
In order to characterize percolation on these hypergraphs we consider their corresponding factor graphs. In particular we indicate with $\hat{S}_m$ the probability that by following a link of a node in layer $m$ we reach a $m$-factor node  ($m$-hyperedge) that belongs to the giant component.
Moreover with $S_m$ we indicate the probability that following a link of a  $m$-factor node ($m$-hyperedge) in layer $m$ we reach a node in the giant component.
Since the corresponding multiplex factor graph of the random multiplex hypergraph is locally tree-like the probabilities $\hat{S}_m$ and $S_m$ can be find to satisfy the self-consistent equations
\bea
	\hat{S}_m &=& p^{[H]}\left[1-(1-S_m)^{m-1} \right],\nonumber \\ 
	S_m &=& p^{[N]}\sum_{{\bf k}}\frac{k_m}{\langle k_m \rangle}P({\bf k})\left[1-\prod_{m^\prime}(1-\hat{S}_{m^\prime})^{k_{m^\prime}-\delta_{m, m^{\prime}}}.\right] \label{eq:S_multiplex}
\eea
These self-consistent equations have a diagramatic interpretation as shown in Figure $\ref{fig:diagram_random_multiplex}$. In particular $\hat{S}_m$ indicates the probability that a $m$-factor node ($m$-hyperedge) reached by following a link in layer $m$, is not iniitially damaged and it is connected at least to a node in the giant component. Instead ${S}'_m$ indicates the probability that a node reached by following a link in layer $m$, is  not initially damaged and it is connected at least to a factor node (hyperedge) -of any possible cardinality- in the giant component. 

The order parameters for percolation on a random multiplex hypergraph are given by the  expected fraction of nodes $R$ and the expected fraction of  hyperedges $\hat{R}$ in the giant component, given by:
\bea
R &=& p^{[N]} \left[1 - \sum_{\bf k} P({\bf k}) \prod_{m} (1 - \hat{S}_m)^{k_m}\right],\nonumber \\
\hat{R}&=& p^{[H]} \left[1 - \sum_{m}\hat{P}(m)(1-S_m)^m \right].
\label{eq:R_multiplex}
\eea
The Eqs.~(\ref{eq:R_multiplex}) together with the Eqs. (\ref{eq:S_multiplex}) fully determine the percolation process in random multiplex hypergraphs and can be used to study the robustness of these structures as a function of the hyperdegree distribution $P({\bf k})$ and the distribution of cardinality of the hyperedges $\hat{P}(m)$.
In particular they can be used to investigate the effect that correlations between the hyperdegrees on different layers have on the robustness properties of the multiplex hypergraph.

One fundamental measure for characterizing the robustness of a multiplex hypergraph with respect to another hypergraph is without any doubt the characterization of the percolation threshold. Indeed a smaller  node (or hyperedge) percolation threshold implies that an hypergraph can display a giant component also when a larger fraction of nodes (or hyperedges) is removed.

The node and hyperedges critical thresholds $p_c^{[N]}$ and $p_c^{[H]}$ can be obtained by imposing that the largest eigenvalue $\Lambda$ of the Jacobian matrix of  Eqs. (\ref{eq:S_multiplex}) calculated for $\hat{S}_m=S_m=0$ is one, i.e. 
\bea
\Lambda=1.
\eea 
By simple analytical calculations it can be show that $\Lambda$ is also the  maximum eigenvalue  of the matrix ${\bf G}$ of elements
\begin{equation}
	G_{mn}=\left\{\begin{array}{lcc}p^{[H]}p^{[N]}(n-1){\avg{k_n k_m}}/{\avg{k_m}}& \mbox{for} &m \neq n\\ p^{[H]}p^{[N]}(m-1){\Avg{k_m (k_m-1)}}/{\Avg{k_m}} &\mbox{for }&{m = n}\end{array}\right.\nonumber
\end{equation}

In the following sections we will predict the percolation threshold in  important examples of random multiplex hypergraphs and we will characterize the role that correlations among hyperdegree of different layers have on the robustness properties of random multiplex hypergraphs.

\subsection{Percolation threshold is some specific cases}
\subsubsection{Hypergraph with fixed cardinality of hyperedges}
For a single layer multiplex  hypergraph including only hyepredges of cardinality $m$, \ie only including $m$-body interactions, we have  
\bea
\hat{P}(m')=\delta_{m,m'}.
\eea
In this case the matrix ${\bf G}$ reduces to a scalar $G$ given by 
\bea
G=G_{mm}=p^{[N]}p^{[H]}(m-1)\frac{\Avg{k_m (k_{m}-1)}}{\Avg{k_m}}
\eea
Therefore the percolation thresholds are obtained by imposing $G=1$, giving  
\bea
p^{[N]}p^{[H]}(m-1)\frac{\Avg{k_m (k_{m}-1)}}{\Avg{k_m}}=1.
\eea
It follows that  in this simple case we recover the expression in Eq. \eqref{eq:pc_random_m}  as we should.

\subsubsection{Independent layers with Poisson generalized degree distribution}
A more interesting case in which we can appreciate the multiplex structure of the problem is given by the case in which the hyperdegree distribution of each layer  of the random multiplex hypergraphs is an independent Poisson distribution with layer-dependent average hyperdegree  $z_m$.
In this case the joint hyperdegree distribution $P(\bf k)$ is given by 
\bea
P({\bf k})=\prod_m P_m(k_m),
\label{eq:indep_degrees}
\eea
with 
\bea
	P_m(k_m) = \frac{e^{-z_m}z_m^{k_m}}{k_m!}.
	\label{eq:Poisson_degrees}
\eea
 
By using the well-known expression for the moments of Poisson distribution, 
\bea
	\frac{\Avg{k_n k_m}}{\Avg{k_n}}=z_m, \ \ \  
	\frac{\Avg{k_m(k_m-1)} }{\Avg{k_m} } = z_m
\eea
we obtain that for this random multiplex hypergraph the matrix ${\bf G}$  has elements $G_{mn}$ given by 
\bea
G_{mn}=p^{[H]}_cp^{[N]}_c (m-1)z_m.
\eea
Since the matrix elements  ${ G}_{mn}$ assume the same value of every element of a given row $m$ the rank of ${\bf G}$ is equal to one, \ie $\mbox{rank}({\bf G})=1$. This implies that  the only non-zero eigenvalue $\Lambda$ of ${\bf G}$ equals the trace of this matrix:
\bea
	\Lambda = \mbox{Tr}({\bf G})=p_c^{[N]}p_c^{[H]}\sum_{m}(m-1)z_m.
\eea
By imposing that $\Lambda=1$ we find that the critical thresholds $p_c^{[N]}$ and $p_c^{[H]}$ satisfy 
\bea
	\frac{1}{p_c^{[N]}p_c^{[H]}}=\sum_{m}(m-1)z_m.
	\label{eq:pcs_Poisson}
	\eea
	This equation can be used to elucidate the relation between the percolation thresholds of the Poisson multiplex hypergraph and the percolation threshold of  single layer Poisson hypergraphs constructed by considering only the hyperedges of a given size.
	Indeed Eq. (\ref{eq:pcs_Poisson}) implies that 
	\bea
	\frac{1}{p_c^{[N]}p_c^{[H]}}=\sum_{m}\frac{1}{p_c^{[N,m]}p_c^{[H,m]}},
\eea
where $p_c^{[N,m]}$ and $p_c^{[H,m]}$ with 
\bea
p_c^{[N,m]}p_c^{[H,m]}=[(m-1)z_m]^{-1},
\eea
 indicating the critical node and hyperedge percolation  thresholds of hypergraphs obtained by considering only the $m$-body iterations in layer $m$.
This implies that the product of the percolation threshold $p_c^{[N]}p_c^{[H]}$ for the multiplex hypegraph model is smaller than the corresponding product  of percolation threshold $p_c^{[N,m]}p_c^{[H,m]}$ for each single layer of the multiplex hypergraph. Therefore the multiplex hypergraph is more robust than every of its layers taken in isolation.

\subsubsection{Independent layers with power-law generalized degree distribution}

Another interesting case of random multiplex hypergraph is the one formed by independent layers each one  with power-law generalized hyperdegree distribution.
In this case the joint hyperdegree distribution $P(\bf k)$ is given by 
\bea
P({\bf k})=\prod_m P_m(k_m),
\eea
with 
\bea
	P_m(k_m) = c_m k_m^{-\gamma_m},
\eea
with $\gamma_m>2$  and $c_m$ indicating the normalization constant.
For this random multiplex hypergraph the matrix ${\bf G}$ has elements 
\bea
	G_{mn}=\left\{\begin{array}{lcc}p^{[H]}p^{[N]}(n-1){\avg{k_n}}& \mbox{for} &m \neq n\\ p^{[H]}p^{[N]}(m-1){\Avg{k_m (k_m-1)}}/{\Avg{k_m}} &\mbox{for }&{m = n}\end{array}\right.\nonumber
\eea

Given that $\gamma_m>2$, we have that each layer is sparse, i.e. $\avg{k_n}$ is finite at the  limit $N\to\infty$. However as soon as one layer is associated to a power-law exponent $\gamma_m\in (2,3]$ the second moment  ${\Avg{k_m (k_m-1)}}/{\Avg{k_m}}$ diverges in the large network limit $N\to \infty$. This implies that the trace of ${\bf G}$ diverges as well, indicating that the maximum eigenvalue diverges. It follows that as soon as one layer has a scale-free generalized hyperdegree distribution, \ie as soon as for at least one layer $m$ we have $\gamma_m\in (2,3]$,  then   
\bea
p_c^{[N]}p_c^{[H]}\to 0,
\eea
in the limit $N\to\infty$. This implies that for standard percolation it is enough that one layer is scale-free to significantly improve the robustness of the random multiplex hypergraph.

\subsection{Effect of correlations between generalized hyperdegrees}

Random multiplex hypergraphs are characterized in general by non-trivial correlations between the hyperdegrees of the same nodes. In particular, given  a random multiplex hypergraph we  indicate with $C_{mn}$ the correlation between the hyperdegrees  of  the same node, connected to hyperedges of cardinality $n$ and $m$ respectively, i.e.
\bea
C_{mn}=\Avg{k_nk_m}-\Avg{k_n}\Avg{k_m}.
\eea
In a random multiplex hypergraph  formed by two layers, this correlations can be modified by permutating the labels of the nodes in a given layer leaving  the hyperdegree distribution of the two layers unchanged.
In particular, it is possible to choose the permutation of the replica nodes in such a way that the correlations among the corresponding generalized degrees is maximized or minimized generating Maximally Positive Correlated Multiplex Hypergraphs and  Maximally Negative Correlated Multiplex Hypergraph. This construction follows very closely the construction to build maximally positive and maximally negative correlated multiplex networks proposed in Ref.\cite{min2014network}.
In particular the Maximally Positive Correlated Multiplex Hypergraph (MPCMH) can be obtained by ranking the generalized hyperdegrees of both layers in increasing order and identifying the label of the nodes with the same rank in both layer. On the contrary Maximally Negative Correlated Multiplex Hypegraph (MNCMH) can be obtained by ranking the generalized degree of one layer in increasing order and the one of the other layer in decreasing order, and by identifying the label of the nodes of the same rank. If the label of the nodes are assigned randomly we will obtain an Uncorrelated Multiplex Hypergraph  (UMH).
\begin{figure*}
	\includegraphics[width=1.8\columnwidth]{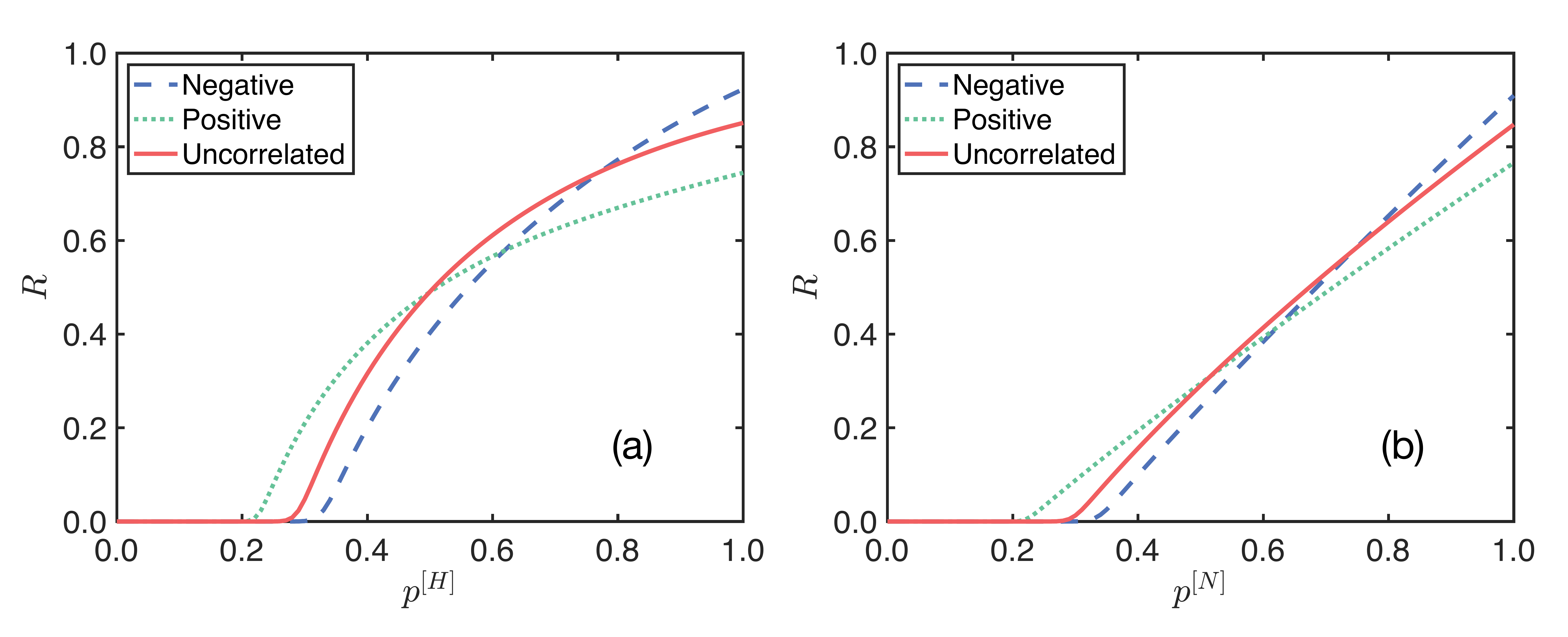} 
	\caption{The fraction $R$ of nodes in the giant component for MPCMH (Positive correlations), for the UMH (Uncorrelated) and  for MNCMH (Negative correlations) is shown for hyperedge percolation (panel(a)) and for node percolation (panel (b)). The considered duplex hypergraph has  $N=10^4$ nodes and  hyperedges of cardinality $m_1=2$ (layer 1) and $m_2=3$ (layer 2). The generalized hyperdegree distributions are  Poisson  with $z_2=0.5$ (for layer 1), $z_3=1.5$ (for layer 2).}
	\label{fig:hyper_fig2}
\end{figure*}
In order  to assess the effect of correlations in the robustness of random multiplex hypergraph here we   focus on a duplex hypergraph and we   investigate the dependence of the percolations thresholds with the correlations coefficient  between the generalized hyperdegrees of the two layers.
In the considered case of a duplex hypergraph with two layers formed by hyperedges of cardinality $m_1$ and $m_2$, the matrix ${\bf G}$ is given by 
\bea
	G = p^{[H]}p^{[N]}\left(
	\begin{matrix}
	\hat{m_1}\kappa_1 & \hat{m_2}\mathcal{K}_1 \\
	\hat{m_1}\mathcal{K}_2 & \hat{m_2}\kappa_2 \\
	\end{matrix}
	\right),
\eea
where $\hat{m}_r=m_r-1$ for  $r\in \{1,2\}$ and where we have used the notation 
\bea
	\frac{\avg{k_{m_r}(k_{m_r}-1)}}{\avg{k_{m_r}}} = \kappa_r, \ \ 
	\frac{\avg{k_{m_1}k_{m_2}}}{\avg{k_{m_r}}} = \mathcal{K}_r,
\eea
for  $r\in \{1,2\}$.
The percolation threshold can be found by imposing that the maximum eigenvalue of the matrix ${\bf G}$ equals one, obtaining
\bea
	p_c^{[N]}p_c^{[H]} = {2}\left[\kappa_1 \hat{m}_1+\kappa_2 \hat{m}_2+\sqrt{\Delta} \right]^{-1}.
	\label{eq:pc_multiplex_correlated}
\eea
with 
\bea
\Delta=\left(\kappa_1 \hat{m}_1-\kappa_2 \hat{m}_2\right)^2+4\mathcal{K}_1\mathcal{K}_2 \hat{m}_1 \hat{m}_2
\eea

We observe that for this percolation problem, the node percolation threshold $p_c^{[N]}$ obtained  when we impose $p^{[H]}=p_c^{[H]}=1$ and the hyperedge percolation threshold $p_c^{[H]}$ obtained when we impose $p^{[N]}=p_c^{[N]}=1$ take the same value. Since the product $\mathcal{K}_1\mathcal{K}_2$ depends on the correlation coefficient $C_{m_1m_2}$ through
\bea
	\mathcal{K}_1\mathcal{K}_2 = \frac{\left(\mathcal{C}_{m_1 m_2}+\avg{k_{m_1}}\avg{k_{m_2}}\right)^2}{\avg{k_{m_1}}\avg{k_{m_2}}},
\eea
Eq. (\ref{eq:pc_multiplex_correlated}) reveals that positive correlations increase the robustness of the multiplex hypergraph against random attack while negative correlations decrease the robustness of the multiplex hypergraph.
\begin{figure*}
	\includegraphics[width=1.9\columnwidth]{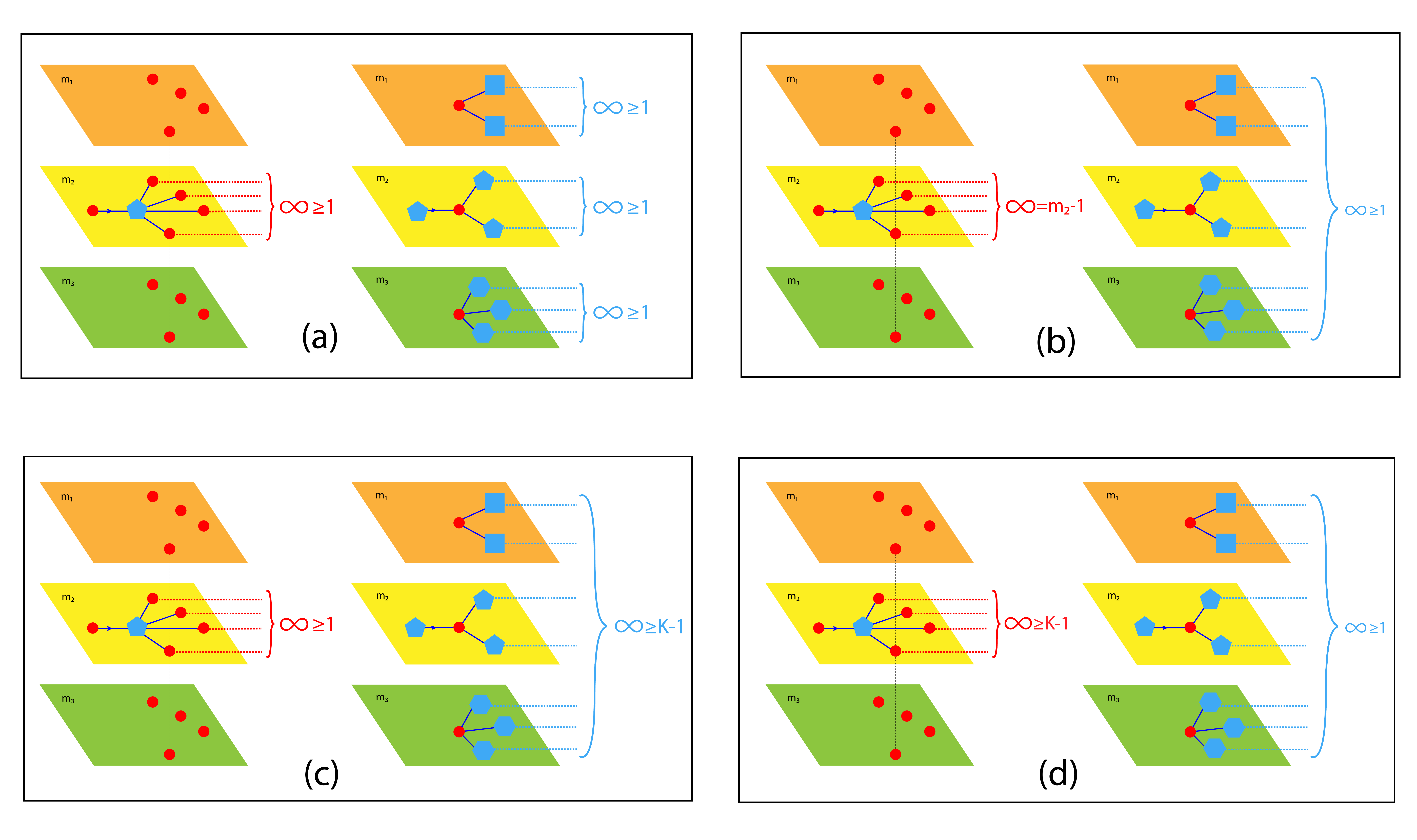} 
	\caption{ Schematic representation of the equations for $\hat{S}_m$ and for $S_m$ determining higher-order percolation models defined on multiplex hypergraphs.
	 Panel (a) represent node interdependent percolation.  Panel (b) represent  hyperedge interdependent percolation.  Pane (c) represent Node $K$-core percolation.  Panel (d) represent hyperedge $K$-core percolation.}\label{fig:summary}
\end{figure*}
In Figure \ref{fig:hyper_fig2} we show the effect of degree correlations on the robustness of random multiplex hypergraph by investigating separately node percolation and hyperedge percolation. We observe that the percolation threshold for node and hyperedge percolation are the same and are in perfect agreement with the analytical results indicating that maximally correlated multiplex hypergraphs have a lower percolation threshold than maximally negative correlated multiplex networks.
The investigation of the the order parameter $R$ versus $p^{[N]}$ in node percolation (when $p^{[H]}=1$) versus $p^{[H]}$ in hyperedge percolation  (when $p^{[N]}=1$)  show  that for both types of percolation a notable effect: the crossing of the curves $R$ versus $p$ (with $p=p^{[H]}$ or $p=p^{[N]}$) calculated for the (MPCMH) and or for the (MNCMH).
This implies that for large values of $p$ the negative degree correlations enhance the robustness of the multiplex hypergraphs with respect to the positive correlations.
In order to understand this phenomenon we note that  
close to the percolation threshold the robustness of the multiplex hypergraph is determined by the high degree nodes, that are less prone to damage in presence of positive correlations, leading to a smaller percolation threshold of MPCMH.
On the contrary, for large values of $p$ the robustness of the multiplex hypergraph, quantified by the fraction $R$ of nodes in the giant component, is highly dependent on the low degree nodes.
In particular the role of low degree nodes is more pronounced when in each layer there is a non neglibile number of isolated nodes.
 In presence of positive  correlations among the  generalized hyperdegrees, the number of nodes isolated in both layers or  connected to a small number of hyperdeges (regardless of their size)  is larger. As a consequence of this MNCMH have a larger fraction of nodes in the giant component than MPCMH, giving an intuition for explaining the fact that for large value of $p$ the order parameter  $R$ becomes larger for MNCMH than for MPCMH in both for node and hyperdegree percolation in Fig. $\ref{fig:hyper_fig2}$.  This effect remains but it is strongly suppressed in absence of isolated nodes.

\section{Higher-order percolation on  multiplex hypergraphs}

\subsection{The landscape of possible higher-order percolation problems}

The topology of random multiplex hypergraph models allows us to explore a large variety of higher-order percolation problems.
Higher-order percolation problems are characterized by illustrating cooperative phenomena where the probability that a node (or a factor node) is active depends on the presence of two or more active neighbours.
These higher-order percolation problems have a highly non-trivial critical behavior and display hybrid discontinuous transitions, tricritical points, and they can even be characterized by more than one critical point as we will show in the following.
Here we investigate and systematically characterize a large variety of higher-order percolation models that can be defined on multiplex hypergraphs.
Inspired by the parallelism between multiplex hypergraphs and multiplex networks \cite{bianconi2018multilayer,Havlin,Havlin2,son2012percolation}  we can define interlayer node interdependence  in multiplex hypergraphs in which a node is active if it has at least a active neighbour on every layer of the multiplex hypergraph. This higher-order percolation is characterized by a hybrid discontinuous transition which can become a continuous transition at a tricritical point if partial interdependence is considered.
However interlayer node interdependence is not the only interdependent model that can be defined on a multiplex hypergraph. In fact we can also consider interdependence associated to  hyperedges, and assume that an hyperedge is active only when all its nodes are active. This highly non-trivial model display hybrid discontinuous transitions if the hyperedges are all involving more than two nodes. In presence of hyperedges of cardinality two (links) the transition can become continuous at a tricritical point in some cases.  Note that this model is the percolation problem corresponding to the higher-order contagion problem proposed and studied in Refs.\cite{iacopini2019simplicial,landry2020effect}.
Another class of higher-order percolation problems is inspired by $K$-core percolation \cite{dorogovtsev2006k,goltsev2006k}. In the  case  of node $K$-core percolation a node is active if at least $K$ of its hyperedges (of any given cardinality) are active, for hyperedge $K$-core percolation instead an hyperedges is active if at least $K$ of its nodes are active. In either one of these last two models the transition is discontinuous as long as $K>2$ and the distributions $P({\bf k})$ and $\hat{P}(m)$ have finite second moment.
These different higher-order percolation problems are summarized in Figure $\ref{fig:summary}$.

\subsection{Interdependent node percolation}
\subsubsection{General framework}
In analogy to interlayer dependency on multilayer networks \cite{bianconi2018multilayer,Havlin,Havlin2,son2012percolation}, we consider the interlayer dependency on hypergraphs. A node in the hypergraph is active when each of its replica nodes  belongs at least to an active  hyperedge, \ie the node belongs to at least one active hyperedge for each possible value of the hyperedge cardinalities $m$. Therefore the probability $\hat{S}_m$ that starting from a node  we reach a $m$-factor node ($m$-hyperedege) that is active and the probability $S_m$ that starting from a $m$-factor node ($m$-hyperedge) we reach a node that is active follow the recursive equations
\bea
	\hat{S}_m& =& p^{[H]} \left[ 1 - (1-S_m)^{m-1} \right],  \nonumber \\ 
S_m &=& p^{[N]}\sum_{\bf k} \frac{k_m}{\avg{k_m}}P({\bf k})\prod_{m^\prime}\left[1-(1-\hat{S}_{m^\prime})^{k_m^{\prime}-\delta_{m,m^\prime}}\right]. \label{eq:interdependent_S}
\eea
Moreover the order parameters $\hat{R}$ and $R$ indicating the fraction of active hyperedges and the fraction of active nodes respectively are given by 
\bea
	\hat{R}& =& p^{[H]} \sum_{m}\hat{P}(m)\left[ 1 - (1-S_m)^{m} \right], \nonumber \\ 
R &=& p^{[N]}\sum_{\bf k} P({\bf k})\prod_{m^\prime}\left[1-(1-\hat{S}_{m^\prime})^{k_m^{\prime}}\right]. \label{eq:interdependent_R}
\eea
As for interdependent percolation on pairwise multiplex networks, these equations lead to discontinuous (and hybrid) phase transitions. 
Let us indicate with $\Lambda$ the maximum eigenvalue of the Jacobian matrix ${\bf J}$ of the system of Eqs.\eqref{eq:interdependent_S}.
The critical point of the discontinuous transition corresponding to non-zero order parameters $R$ and $\hat{R}$ can be  obtained by solving
\bea
\Lambda = 1
\eea
 together with Eqs.~\eqref{eq:interdependent_S} and Eqs.~\eqref{eq:interdependent_R}.
 \subsubsection{Independent layers}
\begin{figure}
	\includegraphics[width=\columnwidth]{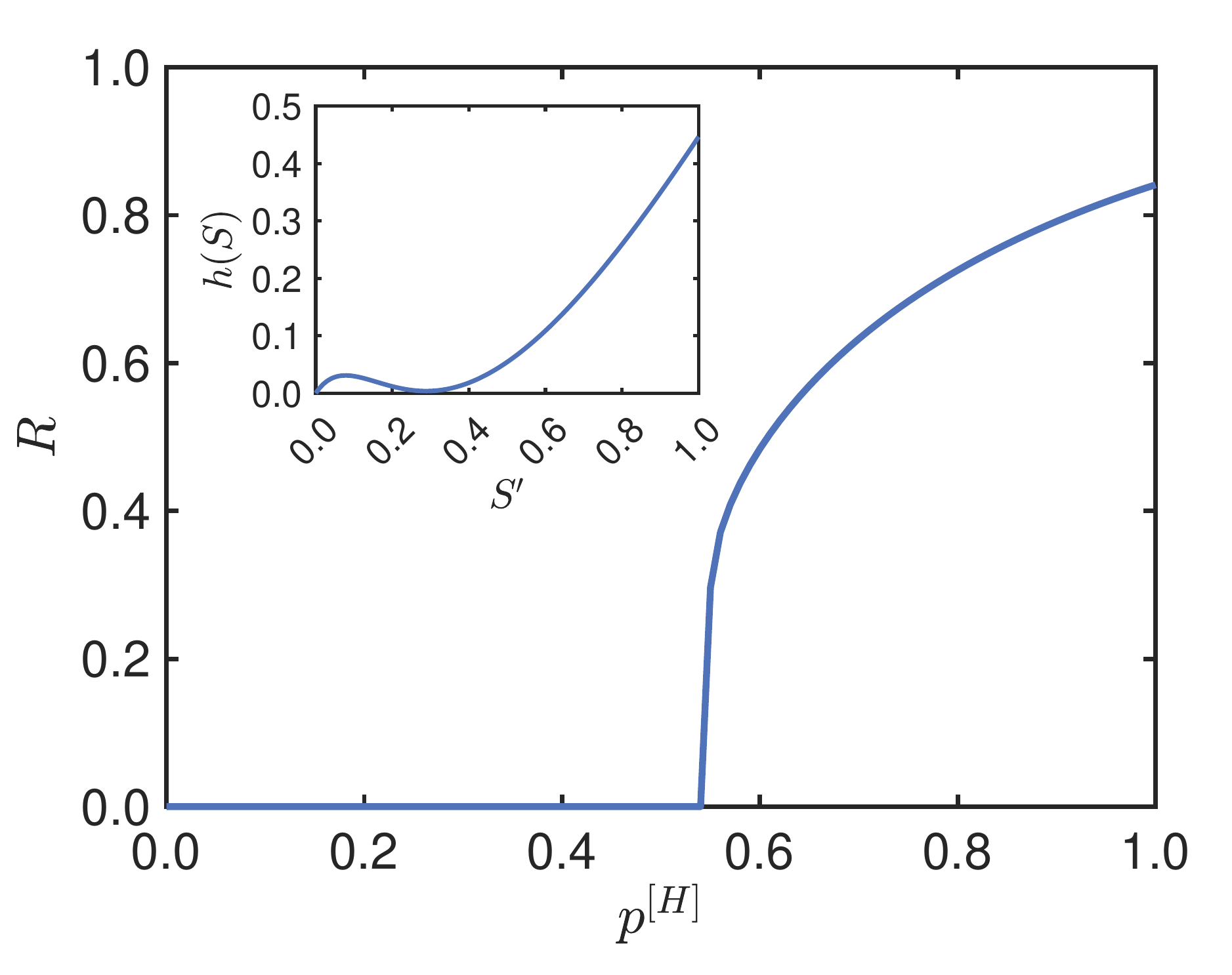}
	\caption{The fraction $R$ of active nodes in interdependent node percolation is shown versus $p^{[H]}$ for a duplex multiplex hypergraph with $p^{[N]}=1$.  The layers of the duplex networks are formed by hyperedges of cardinality $m_1=3$ (layer 1), and $m_2=4$ (layer 2). Both layers have Poisson generalized degree ditribution with $z_3=z_4 = 2.5$. The inset displays the function $h(S)$ defined in Eq. \eqref{eq:h:inter} calculated at the critical point, \ie for $p^{[H]}=p_c^{[H]}$. }\label{fig:interdependent_with_function}
\end{figure}

In order to reveal the mechanism responsible for the discontinuity of the transition, let us consider the model in the simple case in which the generalized degrees of a nodes are independent. In this case the  joint distribution $P({\bf k})$ factorizes according to Eq. \eqref{eq:indep_degrees}. In this limit the equations for $S_m$ and $R$ can be simply written as 
\bea
S_m &=& p^{[N]}\left(1-G_{1,m}(1-\hat{S}_m)\right)\prod_{m^\prime \neq m}\left(1-G_{0,m^\prime}(1-\hat{S}_{m^\prime})\right),\nonumber \\
R &=& p^{[N]}\prod_{m^\prime}\left(1-G_{0,m^\prime}(1-\hat{S}_{m^\prime})\right)
\eea
where $G_{0,m}(x)$ and $G_{1,m}(x)$ indicates the generating functions 
\bea
G_{0,m}(x)&=&\sum_{k_m}P_m(k_m)x^{k_m}, \\
G_{1,m}(x)&=&\sum_{k}\frac{k_m}{\avg{k_m}}P_m(k_m)x^{k_m-1}.
\label{eq:gen}
\eea
By choosing the generalized degree distributions $P_m(k_m)$ to be Poisson and given by Eq. \eqref{eq:Poisson_degrees} these equations further simply as $G_{0,m}(x)=G_{1,m}(x)$. Therefore we have $R=S_m=S$ for every possible value of $R$. Therefore in this simple limit the order parameter $R=S$ obeys a single equations given by 
\bea
h(S)=0,
\label{eq:h}
\eea
with the function $h(S)$ given by 
\bea
h(S)=S-p^{[N]}\prod_{m}\left[1-\exp\left(-p^{[H]}z_m S^{m-1}\right)\right].
\label{eq:h:inter}
\eea
For any node-interdependent multiplex hypergraph with more than one layer this equation describes a discontinuous hybrid phase transition.  Indeed the function $h(S)$ can display a minimum, when this minimum is achieved for $S=S_c$ such that $h(S)=0$ we observe the discontinuous phase transition (see Figure $\ref{fig:interdependent_with_function}$. Therefore the  critical point can be found by solving 
\bea
h(S_c)=h^{\prime}(S_c)=0.
\eea

We can  consider the model in which $p^{[N]}=1$ or the model in which $p^{[H]}=1$. In both models the transition is hybrid.

\begin{figure*}
	\includegraphics[width=1.6\columnwidth]{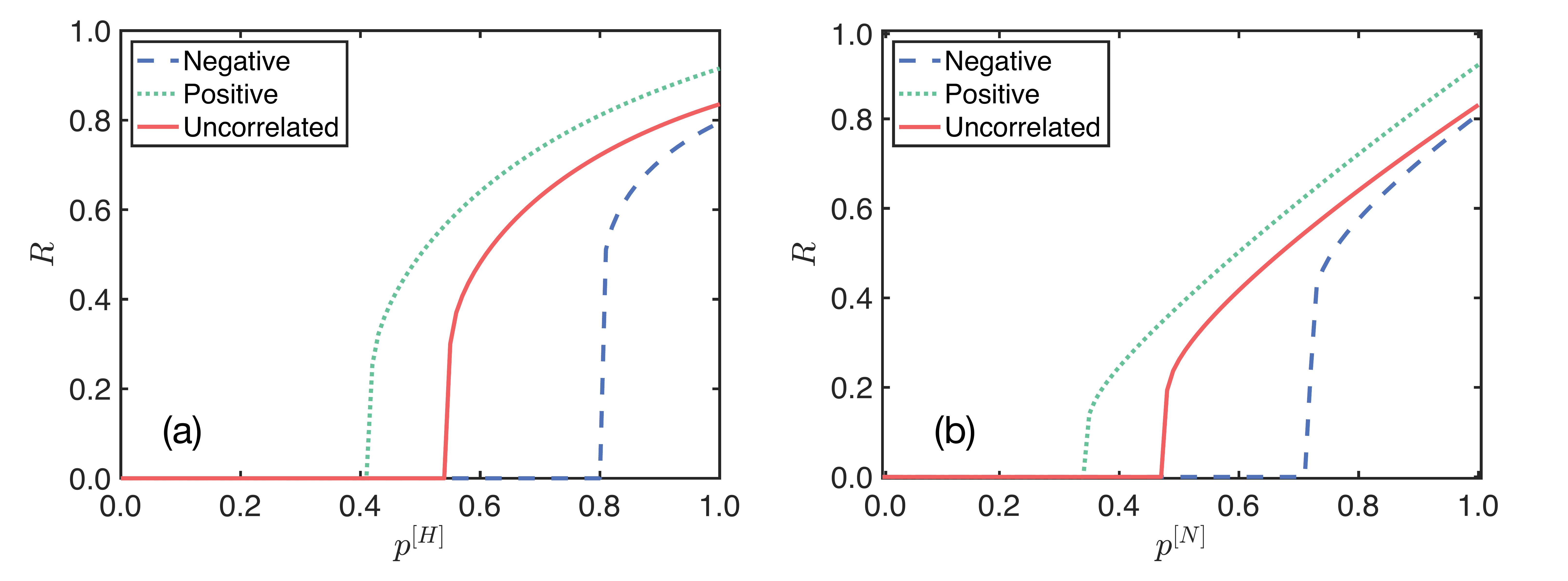}
	\caption{The fraction of active nodes $R$ for interdependent node percolation is plotted versus $p^{[H]}$ when $p^{[N]}=1$ (panel a) and versus $p^{[N]}$ when $p^{[H]}=1$ (panel b) for a MPCMH (Positive correlations) a MNCMH (Negative correlations) and for a UMH (Uncorrelated). The  layers of the duplex hypergraph are formed by hyperedges of cardinality  $m_1=3$ (layer 1), $m_2=4$ (layer 2), with  Poisson layers of average generalized degree $z_3=2.5$,  $z_4=2.5$.	
	}\label{fig:interdependent_correlation}
\end{figure*}
\subsubsection{Effect of correlations between generalized degrees}

Interdependent multiplex hypergraphs display a  higher-order percolation transition that is significantly affected by the correlations between generalized degrees of different layers.
This phenomenon is the higher-order version of the corresponding phenomenon known to occur on pairwise multiplex networks \cite{bianconi2018multilayer,min2014network}. 
By considering a duplex hypergraph with tunable correlations of the generalized degrees of the two layers we observe that MPCMH are more robust than MNCMH, i.e. positive correlations between generalized degrees of different layer increase the robustness of the multiplex hypergraph. This beneficial effect of positive correlations affects the critical threshold of the higher-order percolation model, which is lower for MPCMH than for MNCMH with the same hyperdegree distributions in each of the two layers (see Figure $\ref{fig:interdependent_correlation}$). Interestingly for interdepent multiplex networks the beneficial effect of positive correlations remains effective for every entity of the damage. In fact for this percolation problem, we have that also for  large values of $p^{[H]}$ and $p^{[N]}$ the order parameter $R$ for MPCMH remains always larger that the order parameter $R$ for MNCMH.
This phenomenology differs from the one observed for standard percolation (see Figure $\ref{fig:hyper_fig2}$).
The reason for this different behavior of interdependent percolation is simple: when most of the nodes and most of the links are not initially damaged, the fraction of active nodes is maximized for positive correlations. This remains true also in presence of isolated nodes.
 In fact negative correlations will imply the maximization of nodes which are isolated  on at least one layer, and thus inactive, while positive correlations will minimize the number of nodes isolated in at least one layer.
This simple explanation reveals why in Figure $\ref{fig:interdependent_correlation}$ the order parameter $R$ for MPCMP is always larger that the order parameter $R$ for NPCMP, while we observe a crossing of the two curves for standard percolation (see Figure \ref{fig:hyper_fig2}).

\begin{figure}
	\includegraphics[width=\columnwidth]{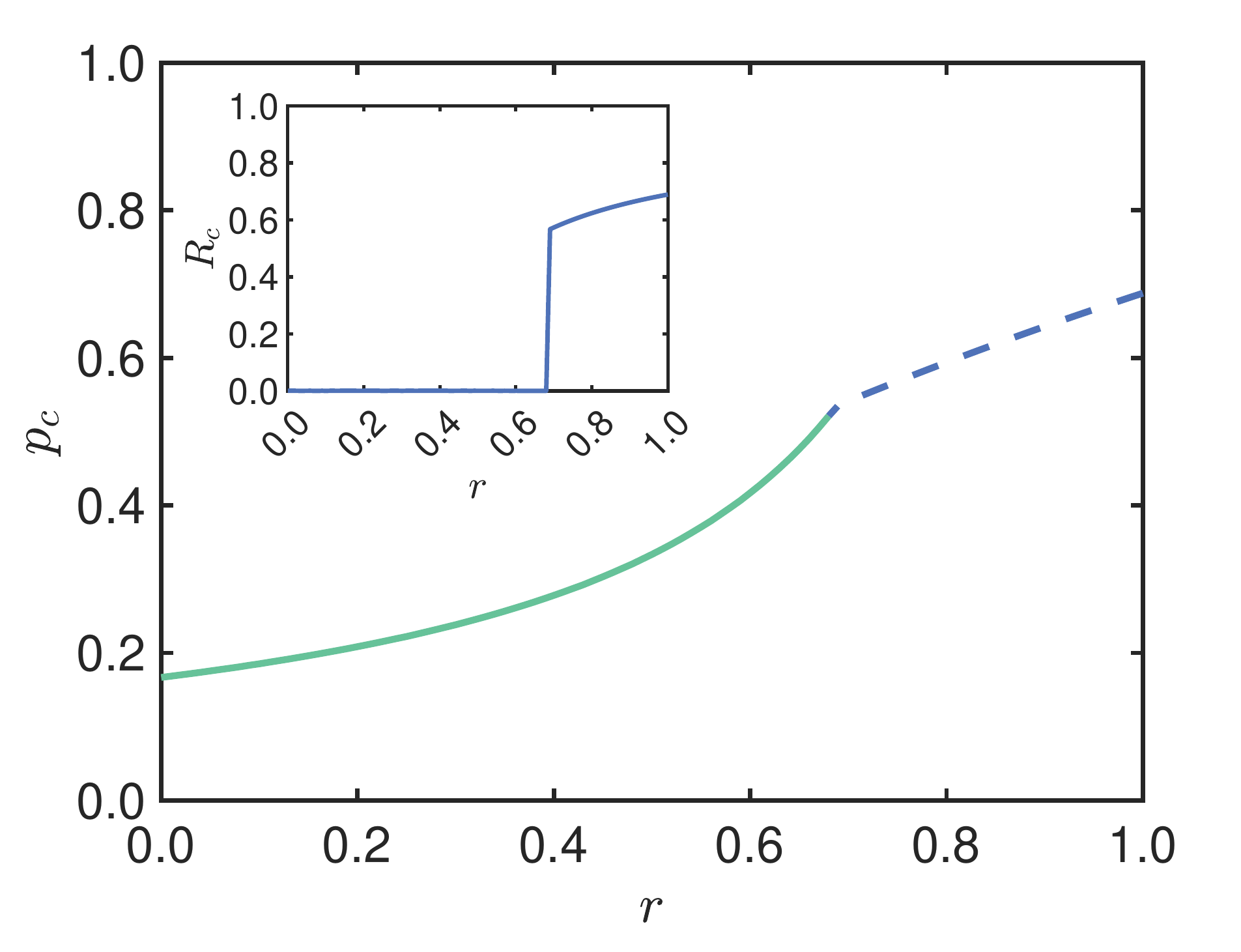}
	\caption{The percolation threshold $p_c=p_c^{[H]}$ of a duplex multiplex hypergraph is plotted versus $r$ for the interdependent node percolation process  with partial interdependence. Solid line correspond to the line of continuous critical point, the dashed line corresponds to the line of discontinuous, hybrid transitions. The tricritcal point separating the two lines is obtained for $r=r_T=0.68\ldots$.  The inset displays the  value $R=R_c$ of the fraction of active nodes at the critical point as a function of $r$ showing that $R_c>0$ for $r>r_T$ indicating that the transition is discontinuous. The  layers of the duplex hypergraph are formed by hyperedges of cardinality  $m_1=3$ (layer 1), $m_2=4$ (layer 2), with  Poisson layers of average generalized degree $z_3=2$,  $z_4=2$. Here $p^{[N]}$ is set equal to one, \ie $p^{[N]}=1$. }\label{fig:tricritical_interdependency}
\end{figure}

\subsubsection{Partial interdependence}

While node-interdependency always leads to discontinuous and hybrid transitions, if  partial interdependence is taken into account it is possible to observe a change of behavior at a triciritical point separating a phase in which the percolation process displays  discontinuous hybrid transitions from a phase in which the process displays continuous transitions.
Partial interdependence has been introduced and investigated in detail for pairwise multiplex networks  \cite{bianconi2018multilayer,Havlin2,son2012percolation}. Here we extend this notion to multiplex hypegraphs highlighting the similarities and differences between the two models.
 By partial interdependence we mean that the interdependence is not always present between the replica nodes but replica nodes are interdependent only with  probability $r$. Therefore for  $r=1$ we recover the node interdependent multiplex hypergraph studied in the previous paragraph and displaying a discontinuous hybrid transition, while for $r=0$ we recover the standard percolation model studied in Sec. III displaying a continuous transition. 
Let us restrict our discussion here to the simple case of independent generalized degrees with join generalize degree distribution $P({\bf k})$ given by Eq. \eqref{eq:indep_degrees}.
In this case the equation for $\hat{S}_m$ and the equation for $\hat{R}$ remains unchanged (given by the first of Eqs. \eqref{eq:interdependent_S} and \eqref{eq:interdependent_R}, however the equations for $S_m$ and $R$ change and are given by 
\bea
S_m = p^{[N]}\left(1-G_{1,m}(1-\hat{S}_m)\right)\prod_{m^\prime \neq m}\left(1-rG_{0,m^\prime}(1-\hat{S}_{m^\prime})\right),\nonumber  \\
R = p^{[N]}\left(1-G_{0,m}(1-\hat{S}_m)\right)\prod_{m^\prime \neq m}\left(1-rG_{0,m^\prime}(1-\hat{S}_{m^\prime})\right).\nonumber
\eea
Interestingly due to the higher-order nature of the multiplex hypergraphs these equations cannot be reduced to a single equation in the case of Poisson layers with generalized degree distribution given by Eq. \eqref{eq:Poisson_degrees}.
However the phase diagram of the model can be investigated numerically. The phase diagram is characterized by a tricritical point separating a regime with $r<r_T$ for which we observe continuous transitions and a regime with  $r>r_T$ in which we observe a discontinuous hybrid phase transition. 
Let us consider the case in which either nodes ($p^{[H]}=1,p^{[N]}=p$) or hyperedges ($p^{[N]}=1,p^{[H]}=p$) are randomly removed with probability $1-p$.
In this case the tricritical point $(r_T,p_T)$ can be found   numerically by solving the self-consistent equations for  $\hat{S}_m$ and $S_m$ together with 
\bea
\Lambda=1
\label{eq:tricritical_jacobian}
\eea 
where $\Lambda$ is the largest eigenvalue of the Jacobian matrix of the equations determining  $\hat{S}_m$ and $S_m$  (see Fig.\ref{fig:tricritical_interdependency}).

\begin{figure*}
	\includegraphics[width=2\columnwidth]{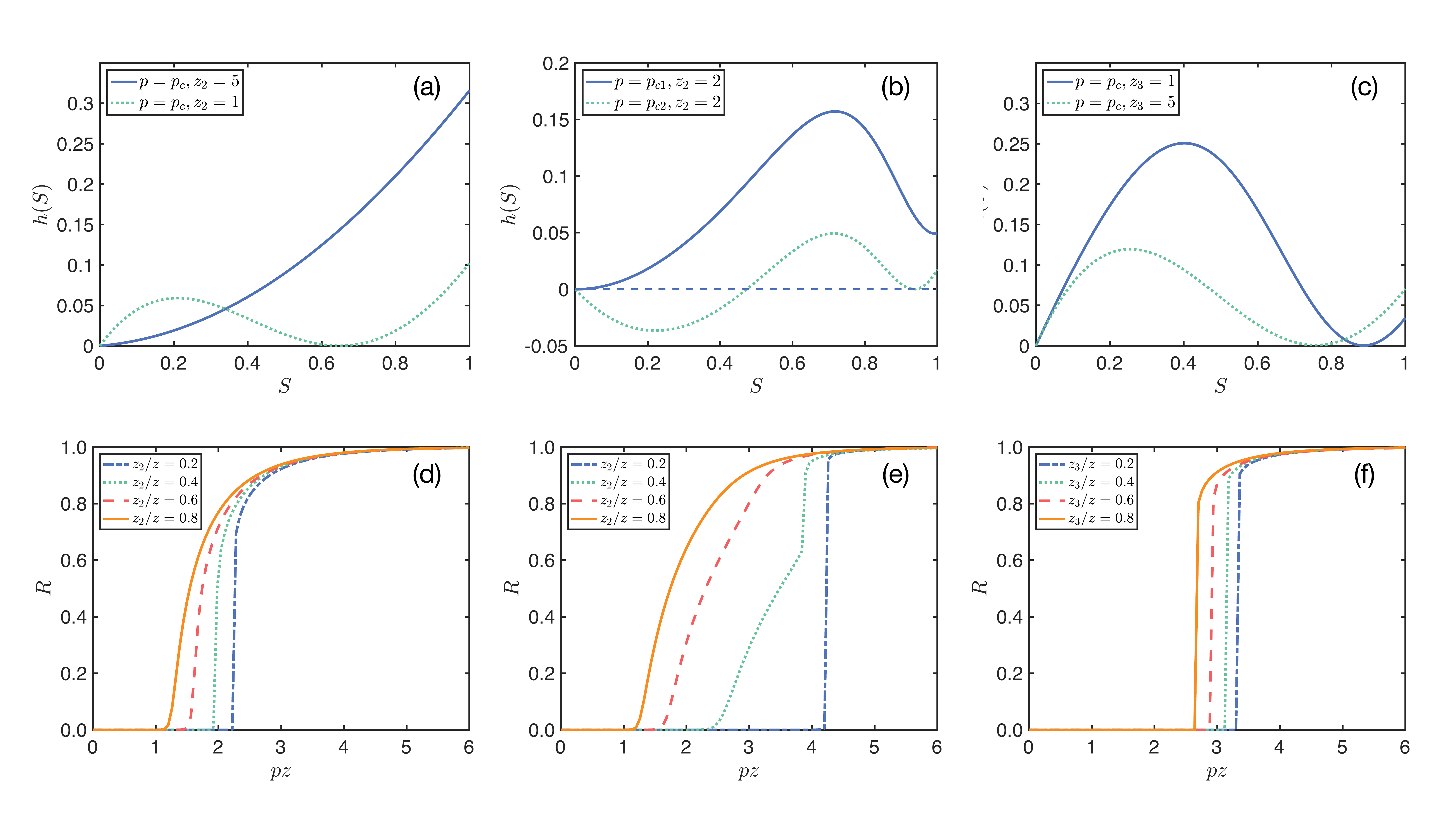}
	\caption{The critical behavior of the interdependent hyperedge percolation process on a duplex hypergraph is investigated by plotting the function $h(S)$ defined in Eq. \eqref{eq:h} versus $S$  (panels (a) and (c)) and by displaying the the fraction of active nodes  $R$  for different values of $p=p^{[H]}$ (panels (d) and (f)).
	The duplex hypergraphs have with layers having hyperedges of cardinality  $m_1=2$, $m_2=3$ (panel (a) and (d)), $m_1=2$, $m_2=10$ (panel (b) and (e)), and $m_1=3$, $m_2=5$ (panel (c) and (f)). Each layer is characterized by Poisson hyperdegree distributions with average degree $z_{m_1}$ (layer 1) and $z_{m_2}$ (layer 2) with $z_{m_1}+z_{m_2}=z=6$.  In panel (d) we observe a continuous transitions and a discontinuous transitions occurring for different values of   $z_2$. In panel (e) we observe that the model can display, for the same value of $z_3$, two critical points $p_{c1}$ and $p_{c2}$ corresponding to a continuous and discontinuous transition occurring at a non zero value of the order parameter .In panel (f) we show that all the transitions are discontinous.}\label{fig:fullkcore}
\end{figure*}

\subsection{Interdependent hyperedge percolation}
\subsubsection{General framework}
 
 In this section we introduce the higher-order interdependent hyperedge percolation model. In this model an hyperedge is active only if all its nodes are active as well and a node is active if at least one of its hyperedges is active.
This model is here chosen because of its complementarity with the node-interdependence where a node is active if all its replica nodes are active, \ie all its replica nodes belong to at least an active hyperedge.
Interestingly the independent hyperedge percolation problem can be related to the model of higher-order social contagion  proposed in Ref. \cite{iacopini2019simplicial} and investigated on random hypergraphs in Ref. \cite{landry2020effect}. Indeed in higher-order contagion model a node is infected if at least one of its $m$-hyperedges connects the new node to a set of $m-1$ infected nodes.
The inderdependent hyperedge percolation model and the higher-order contagion model can be mapped to each other.  However there is a significant difference  between the contagion model on hypergraphs and interdependent hyperedge percolation. While the contagion process admits a region of bistability in which the number of infected nodes can acquire either a larger or a  smaller value depending on the initial conditions of the dynamics, in the corresponding region of the phase diagram, the interdependent hyperedge percolation does not display bistability. Indeed, although the self-consistent equation for the order parameter admits two solutions, the order parameter $R$ always takes the value of the  largest solution of the self-consistent equations. Broadly speaking percolation can be seen as an optimization problem in which one characterizes the maximum number of nodes that are connected under the condition imposed by the combinatorics of the process.
 
In the hyperedge interdepent percolation model the probability $\hat{S}_m$ that starting from a random node we reach an  $m$-factor node ($m$-hyperedge)  which is active, and the probability $S_m$ that starting from a $m$-factor node ($m$-hyperedge) we reach a node that is active are given by 
\bea
	\hat{S}_m &=& p^{[H]} S_m^{m-1}, \nonumber \\
	S_m &=& p^{[N]}\sum_{\bf k}\frac{k_m}{\langle k_m \rangle}P({\bf k})\left[1-\prod_{m^\prime}(1-\hat{S}_{m^\prime})^{k_{m^\prime}-\delta_{m, m^{\prime}}}\right]. 
	\label{eq:interdependent_hyperedge_H}
\eea
Moroever the order parameter $\hat{R}$ and $R$ indicating the fraction of active hyperedges and active nodes respectively are given by 
\bea
	\hat{R} &=& p^{[H]}\sum_{m}\hat{P}(m) S_m^{m}, \nonumber \\
	R &=& p^{[N]}\sum_{\bf k}\frac{k_m}{\langle k_m \rangle}P({\bf k})\left[1-\prod_{m^\prime}(1-\hat{S}_{m^\prime})^{k_{m^\prime}}\right]. \label{eq:interdependent_hyperedge_R}
\eea
These equations differ with respect to the equation valid for standard percolation. In particular the equations for $\hat{S}_m$ and $\hat{R}$  imply that an hyperedge can be active only if all its nodes are also active.
Therefore we note that is the multiplex hypergraph contains only one layer and the layer capture only pairwise interactions, i.e. $m_1=m=2$ then this model reduces to standard percolation, however as long as the hypegraph contains hyperedges of cardinality $m\neq 2$ the interdependent hyperedge percolation problem differs from standard percolation. 
In the following paragraphs we will investigate the nature of the percolation transition and the effect of correlations among generalized degrees observed for this model.
\begin{figure*}[htb!]
	\includegraphics[width=1.8\columnwidth]{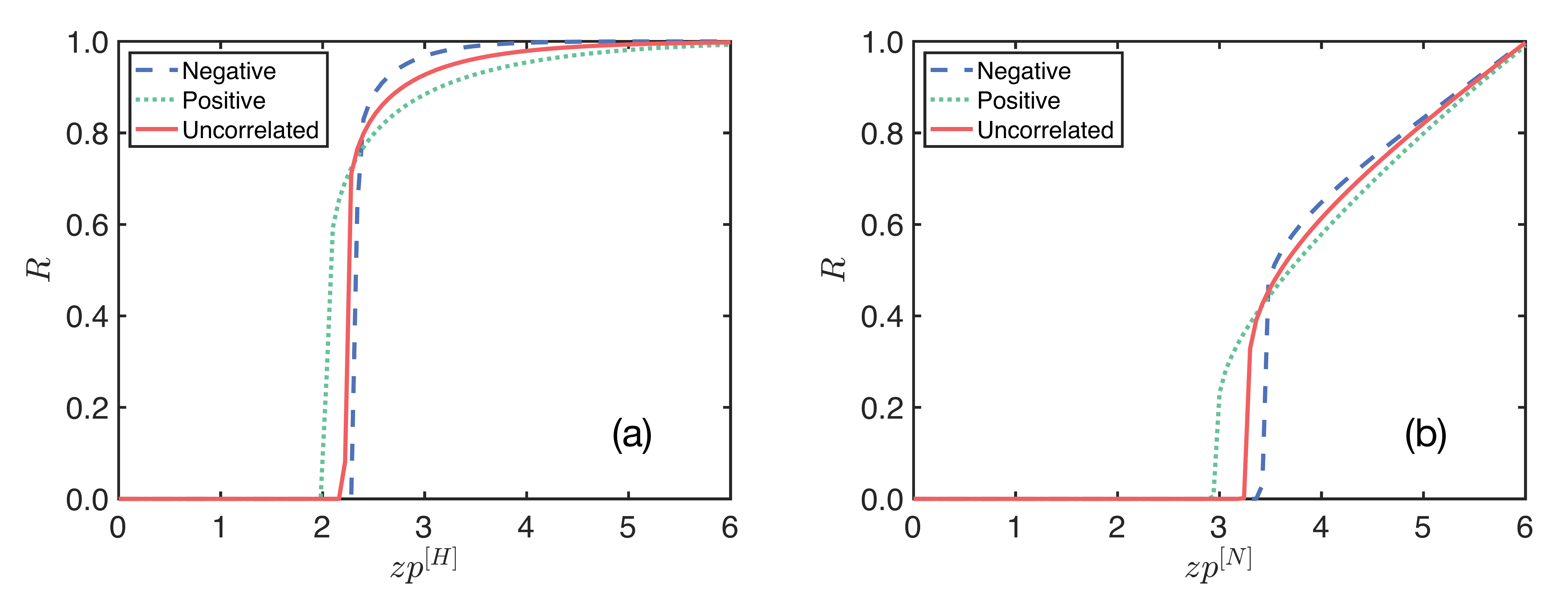}
	\caption{The fraction $R$ of active nodes $R$ in the interdependent hyperedge percolation is plotted versus $p^{[H]}$ when $p^{[N]}=1$ (panel a) and versus $p^{[N]}$ when $p^{[H]}=1$(panel b) for a MPCMH (Positive correlations) a MNCMH (Negative correlations) and for a UMH (Uncorrelated). The  layers of the duplex hypergraph are formed by hyperedges of cardinality  $m_1=2$ (layer 1), $m_2=3$ (layer 2), with  Poisson layers of average generalized degree $z_2=4.8$,  $z_3=1.2$ and $z=z_2+z_3=6$. }
	\label{fig:hyperedge_correlations}
\end{figure*}
\subsubsection{Independent layers}
In is instructive to investigate the critical properties of hyperedge interdependence for a multiplex hypergraph with independent layers. In this case the Eqs. \eqref{eq:interdependent_hyperedge_H}  and Eqs. \eqref{eq:interdependent_hyperedge_R} reduce to 
\bea
\hat{S}_m&=&p^{[H]}S_m^{m-1},\nonumber\\
S_m&=&p^{[N]}\left[1-G_{1,m,}(1-\hat{S}_m)\prod_{m^{\prime}\neq m}G_{0,m^{\prime}}(1-\hat{S}_{m^\prime}),\right]\nonumber \\
R&=&p^{[N]}\left[1-\prod_{ m}G_{0,m,}(1-p^{[H]}S_m^{m-1})\right],\nonumber \\
\hat{R}&=&=p^{[H]}\sum_{m}\hat{P}(m)S_m^{m},
\eea
where the generating functions $G_{0,m}(x)$ and $G_{1,m,}(x)$ are defined in Eq. (\ref{eq:gen}).
By considering Poisson layers with generalized degree distribution given by Eq. \eqref{eq:Poisson_degrees} we observe that $S_m=R=S$ for every value of $m$ with $S$ satisfying 
\bea
S&=&p^{[N]}\left[1-\exp\left(-p^{[H]}\sum_{m}z_m S^{m-1})\right)\right]\nonumber \\
\eea
In the case of 2-layer multiplex hypergraph we obtain that $S$ satisfies 
\bea
h(S)=0,
\eea
with
\bea
\hspace*{-3mm}h(S)=S - p^{[N]}\left\{1-\exp\left[-p\left(z_{m_1} S^{m_1-1} + z_{m_2} S^{m_2-1}\right)\right]\right\}.
\eea
Let us  fix the expected number of hyperedges incident to a node, regardless of their cardinality, by imposing
\bea
z_{m_1} + z_{m_2} = z,
\eea

and let us investigate the  nature of the interdependent hyperdge percolation transition as a function of $z_{m_1}$.
Let us start with the specific example of having two layers with $m_1=2$ and $m_2=3$. If $z_2=z$ and $z_3=0$ the multiplex hypergraph reduces to a single network, and the transition is the standard percolation transition, which occurs at a critical point characterized by satisfying 
\bea
h(0)=h^{\prime}(0)=0.
\eea
In the other extreme case in which $z_2=0$ and $z_3=z$, the multiplex hypergraph reduces to a single layer hypergraphs including only $3$-hyperedges. In this case the transtion is discontinuous and is obtained at a non zero value $S=S_c$ for which 
\bea
h(S_c)=h^{\prime}(S_c)=0.
\eea
These are the two limiting cases of a region of the phase space in which we observe a continuous transition and of a region of phase space in which we observe a discontinuous transition.
These two regions are separated by a tricritical point observed at the value of $z_{m_1}=z_{T}$ that satisfies 
\bea
h(0)=h^{\prime}(0)=h^{\prime\prime}(0)=0.
\eea
For hyperedge interdependent percolation with $p^{[N]}=1$ we obtain the triciritcal point at 
\bea
z_{T}&=&\frac{2}{3}z,\\ \nonumber 
p^{[H]}_T &=& \frac{3}{2z}.
\eea
For hyperedge interdependent percolation  with $p^{[H]}=1$ we obtain the tricritical point at 
\bea
z_{T}&=&\sqrt{1+2z}-1,\\ \nonumber
p^{[N]}_T &=& \frac{\sqrt{1+2z}+1}{2z}. 
\eea
As we change the values of $m_1$ and $m_2$ characterizing the two layers of the duplex multiplex network different scenarios emerges.
For $m_1>2$ and $m_2>2$, the transition is always discontinuous.
 Interestingly, as is shown in Figure \ref{fig:fullkcore}(b), when $m_1=2$ and $m_2>3$ the hyperedge interdependent percolation can display not just one but also two percolation transition. The first transition describes the emergence of the generalized giant component and is continuous, the second transition indicates a discontinuity of the order parameter $R$ from a non-zero value to another non-zero value.
As far as we know this phenomenon has not been reported before, not even for the higher-order contagion model studied in Refs. \cite{iacopini2019simplicial,landry2020effect} but can have an interesting interpretation in that context as a sudden activation of hyperedges of larger cardinality.

\subsubsection{Effect of correlations}

The general equations determining hyperedge interdependent percolation can be also used to study the effect of correlations between the generalized degrees of the replica nodes in different layers. In this case, regardless the nature of the phase transition, we observe that MPCHM display a  transition threshold smaller than MNCMH, indicating that the system is able to sustain more damage. However for small entity of the damage, and in the extreme case in which the multiplex hypergraph is not damaged, the MPCMH have a smaller giant component than the MNCMH. This phenomenon is expected as it has the same explanation of the corresponding phenomena observed and discussed in Sec. III for  the case of standard percolation (see Figure \ref{fig:hyperedge_correlations}).

\subsection{Node $K$-core percolation}
In this section we propose the  $K$-core node percolation on random multiplex hypergraphs. This model is a higher-order percolation process that generalizes $K$-core percolation of single pairwise networks to the multiplex hypergraphs. 
In $K$-core node percolation a node is active if has at least  $K$ active neighbours. In  $K$-core node percolation defined on a multiplex hypergraph, a node is active if it belongs at least to $K$ hyperedges regardless of their cardinality.
 The probability $\hat{S}_m$ that starting from a node  we reach an $m$-factor node ($m$-hyperedege) that is active and the probability $S_m$ that starting from a $m$-factor node ($m$-hyperedge) we reach a node that is active follow the recursive equations
\bea
\hat{S}_m&=&p^{[H]}\left[ 1-(1-S_m)^{m-1} \right], \nonumber
\\
	S_m &=&p^{[N]}\sum_{{\bf k}}^{\ \prime} \frac{k_m}{\avg{k_m}} P({\bf k}) \left[ 1-\sum_{q=0}^{K-2}B_q({\bf k})\right],
\label{eq:activate_hyperedge_1}
\eea
where 
$\sum_{{\bf k}}^{'}$ indicates the sum over of ${\bf k}$ such 
\bea
	\sum_{m} k_{m} \geq  K.
\eea
Here $B_q({\bf k})$ is given by
\bea
	B_q({\bf k}) = \sum_{\left\{ q_{m^{\prime}} \right\}}^{\ \ \ \  \ \ \prime\prime} \prod_{m^{\prime}}\left[\binom{k_m^{\prime}-\delta_{m,m^{\prime}}}{q_{m^{\prime}}} \hat{S}_{m^{\prime}}^{q_{m^{\prime}}} (1-\hat{S}_{m^{\prime}})^{k_m^{\prime}-\delta_{m,m^{\prime}}-q_{m^{\prime}}}\right],\nonumber 
\eea
where $\sum_{\left\{ q_{m^{\prime}} \right\}}^{''}$ indicates the sum over of $\{q_{m^{\prime}}\}$ such that 
\bea
	\sum_{m^{\prime}} q_{m^{\prime}} = q.
\eea

\begin{figure}
	\includegraphics[width=\columnwidth]{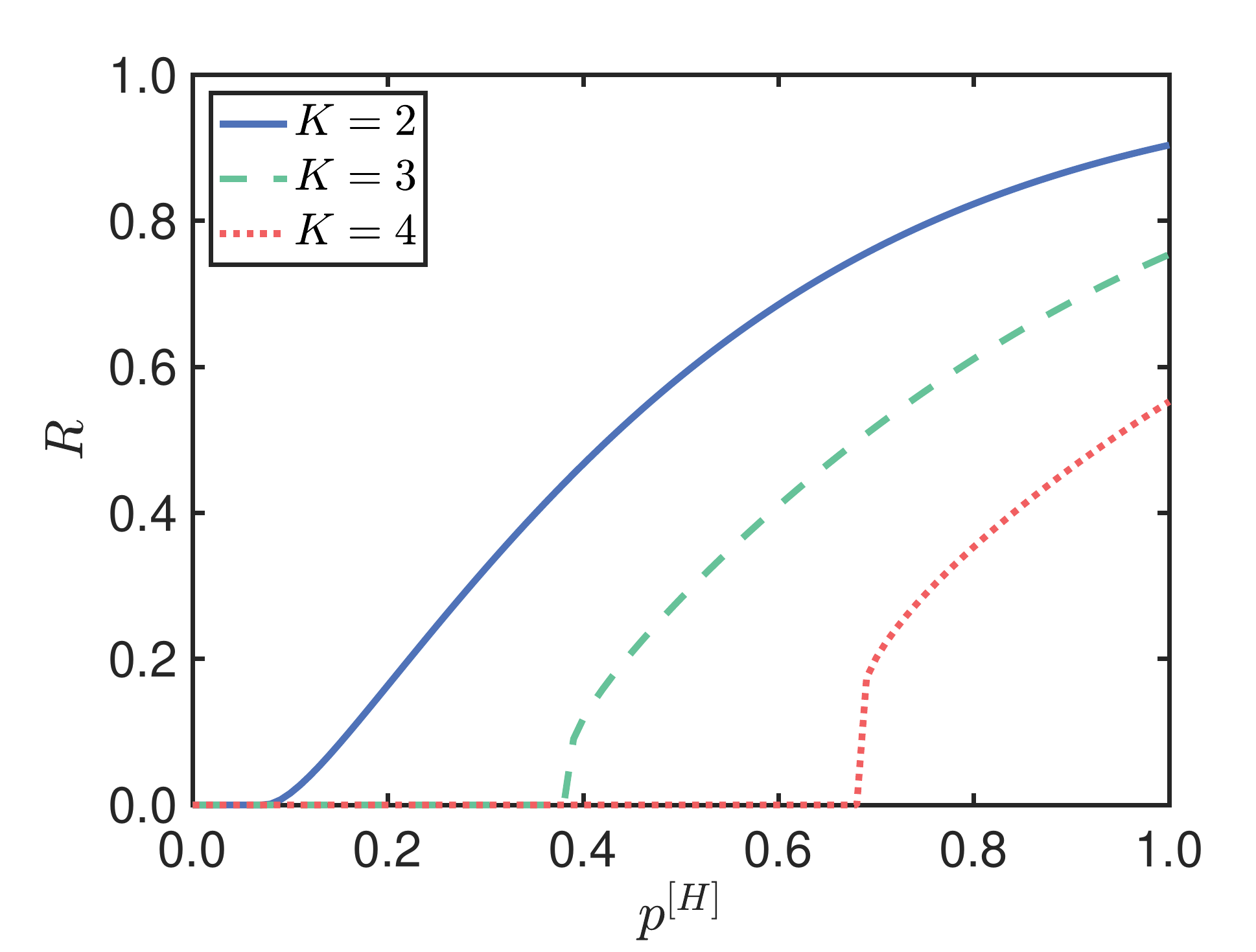}
	\caption{{The fraction $R$ of active nodes for  the node $K$-core percolation on  duplex hypegraphs with independent Poisson layers is shown versus the probability of retaining a hyperedge $p^{[H]}=p$.  The duplex hypergraph includes $N=10^4$ nodes and has  layers  formed by hyperedges of cadinality  $m_1=4$ and $m_2=5$ with independent Poisson generalized hyperdegree distributions  with average $z_4=z_5=2$. Here $p^{[N]}$ is fixed to the constant value $p^{[N]}=1$. The node $K$-core percolation is discontinuous for $K>2$.}}\label{fig:kcore_node}
\end{figure}

The order parameters $R$ and $\hat{R}$ expressing the fraction of nodes ($R$) and the fraction of hyperedges ($\hat{R}$) in the node $K$-core are given by  
\bea
\hat{R}&=&p^{[H]}\left[1-\sum_{m}\hat{P}(m)(1-S)^{m-1}\right],\nonumber \\
R&=&p^{[N]}\left[1-\sum_{{\bf k}}^{\prime}P({\bf k})\sum_{q=0}^{K-1}D_q\right],
\eea
with $D_q$ given by 
\bea
D_q=\sum_{\left\{ q_{m^{\prime}} \right\}}^{\ \ \ \  \ \ \prime\prime} \prod_{m^{\prime}}\left[\binom{k_m^{\prime}}{q_{m^{\prime}}} \hat{S}_{m^{\prime}}^{q_{m^{\prime}}} (1-\hat{S}_{m^{\prime}})^{k_m^{\prime}-q_{m^{\prime}}}\right].
\eea
It follows that the equations for node $K$-core percolation reduce to standard $K$-core percolation if the multiplex hypergraph if formed by a single layers encoding for  hyperedges of cardinality $m=2$ (\ie links).
For node $K$-core percolation, like for node $K$-core percolation on pairwise networks \cite{dorogovtsev2006k,goltsev2006k}, we observe that the percolation transition is discontinuous and hybrid as long as $K>2$ provided that the generalized degree distributions have finite second moment (see Figure $\ref{fig:kcore_node}$).

\subsection{Hyperedge $K$-core percolation}
\begin{figure}
	\includegraphics[width=\columnwidth]{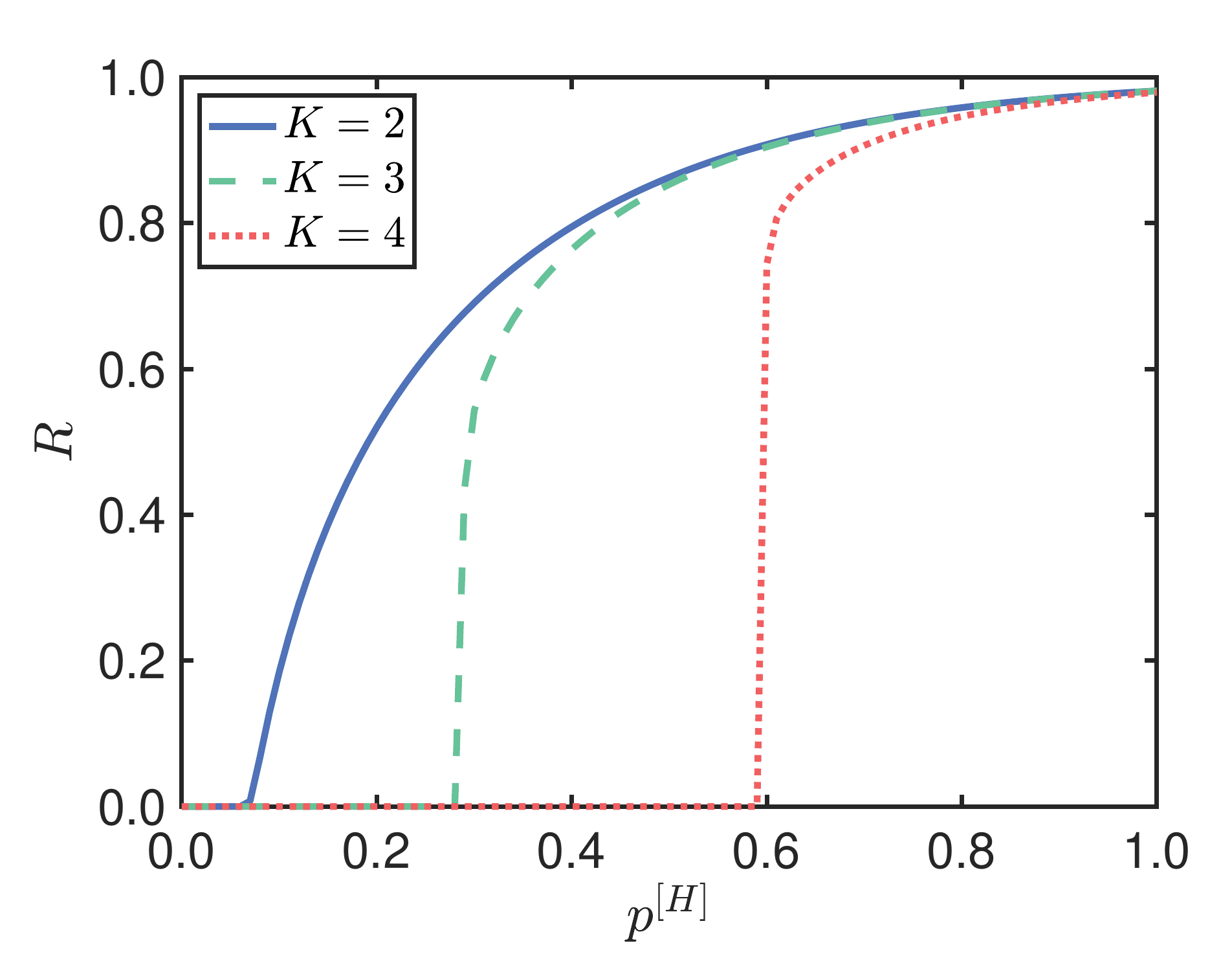}
	\caption{
	The fraction $R$ of active nodes for  the hyperedge $K$-core percolation  on duplex hypegraphs with independent Poisson layers is shown versus the probability of retaining a hyperedge $p^{[H]}=p$.  The duplex hypergraph includes $N=10^4$ nodes and has  layers  formed by hyperedges of cadinality  $m_1=4$ and $m_2=5$ with independent Poisson generalized hyperdegree distributions  with average $z_4=z_5=2$. Here $p^{[N]}$ is fixed to the constant value $p^{[N]}=1$. The transition is discontinuous for $K>2$.}\label{fig:kcore_edge}
\end{figure}

Hyperedge $K$-core percolation is here defined as  a higher-order percolation process occurring on multiplex hypegraphs in which a hyperedge is active only if at least $K$ (with $K\geq 2$) nodes belonging to it are also active.
In this case,  the probability $\hat{S}_m$ that starting from a node we reach a $m$-factor node ($m$-hyperedge)  that is active and the probability $S_m$ that starting from a $m$-factor node ($m$-hyperedge) we reach a node that is active are given by 
\bea
\hat{S}_m &=& \left\{\begin{array}{lcc} p^{[H]}\left[1-\sum_{q=0}^{K-2}\hat{B}_q(m)\right] &\mbox{for} &m\geq K\\
	0  &\mbox{for} & m< K\end{array}\right.,\nonumber \\
	S_m &=& p^{[N]}\left[1-\sum_{{\bf k}}\frac{k_m}{\avg{k_m}} P({\bf k}) \prod_{m^\prime}(1-\hat{S}_{m^\prime})^{k_{m^\prime}-\delta_{m, m^{\prime}}}\right],
	\label{eq:S_hat_prime_kcore}
\eea
where $B_q(m)$ can be expressed as
\bea
\hat{B}_q(m)=\left(\begin{array}{c}{m-1}\\{q}\end{array}\right)(S_m)^q(1-S_m)^{m-1-q}. 
\eea
Similarly we can define the  order parameters $R$ and $\hat{R}$ indicating the fraction of nodes and hyperedge  that are active as
\bea
	\hat{R} &=& p^{[H]}\sum_{m\geq K}\hat{P}(m)\left[1-\sum_{q=0}^{K-1}\left(\begin{array}{c}{m-1}\\{q}\end{array}\right)S_m^q(1-S_m)^{m-q} \right],\nonumber \\
		R &=& p^{[N]}\left[1-\sum_{\bf k}P({\bf k})\prod_{m^\prime}(1-\hat{S}_{m^\prime})^{k_{m^\prime}}\right].
	\label{eq:R_kcore}
\eea
For hyperedge $K$-core percolation like for $K$-core percolation on pairwise networks \cite{dorogovtsev2006k,goltsev2006k}, we observe that the percolation transition is discontinuous and hybrid as long as $K>2$ provided that the distribution $\hat{P}(m)$ has finite second moment (see Figure $\ref{fig:kcore_edge}$).

\section{Conclusions}

In this paper we have provided a comprehensive framework to study standard and higher-order percolation on random multiplex hypergraphs.
Random multiplex hypergraphs are a natural generalization of random hypergraphs where the hyperedges of different cardinality are associated to different layers of the multiplex.  This modelling framework is very comprehensive and is here used to investigate the rich interplay between the topology of  hypergraphs and the properties of standard and higher-order percolation defined on these structures.
We reveal how interlayer correlations among the generalized degree of replica nodes can affect the critical properties of standard percolation. In particular we show that close to the percolation transition positive correlations enhance the robustness of multiplex hypergraphs while when the initial damage is minor negative correlations can be beneficial to network robustness.
We show how the multilayer nature of multiplex hypergraphs can be exploited to define a number of higher-order percolation processes. In particular we propose two models generalizing interdependent percolation in multiplex networks and contagion model in hypergraphs (the interdependent node and the interdependent  hyperedge percolation) and two models generalizing $K$-core percolation to hypergraphs (the node $K$-core and hyperedge $K$-core percolation).
These models are here shown to display a rich phenomenology including discontinuous hybrid phase transitions, tricritical points, and multiplex phase transitions together with non-trivial effects due to the  interlayer correlations among the generalized  degrees.

Although our aim is to provide a  comprehensive view of the possible
higher-order percolation processes on random multiplex hypergraphs we are aware that the processes investigated in this work are not exhausitive of the many relevant percolation processes that can be defined on these structures.
We hope that this work can generate further interest in the interplay between the structure of higher-order networks and their dynamics and that the revealed properties of percolation on multiplex hypegraphs can open new insights also for the study of other dynamical processes such as epidemic spreading and social contagion.

\bibliographystyle{apsrev4-1}
\bibliography{biblio}

\end{document}